\documentclass[12pt,a4paper]{article}

\usepackage{arxiv}

\usepackage[utf8]{inputenc}
\usepackage[english]{babel}
\usepackage[T1]{fontenc}
\usepackage{microtype}
\usepackage{graphicx}
\usepackage{hhline}
\usepackage{pgf}
\usepackage{subcaption}
\usepackage{siunitx}
\usepackage{url}
\usepackage{booktabs}
\usepackage{amsfonts}
\usepackage{amsmath}
\usepackage{bm}
\newcommand*{\B}[1]{\ifmmode\bm{#1}\else\textbf{#1}\fi}
\usepackage{nicefrac,xfrac}
\usepackage[]{authblk}
\usepackage[format=hang]{caption}
\usepackage{multirow, bigdelim}
\usepackage{xcolor}
\usepackage[linkbordercolor={0.818 0.394 0.394},urlbordercolor={0.366 0.528 0.745},hidelinks]{hyperref}
\usepackage{placeins}
\usepackage{import}
\usepackage[printonlyused, nohyperlinks]{acronym}
\usepackage{csquotes}
\usepackage[super]{nth}
\usepackage{blindtext}
\usepackage{float}

\title{Systematic Misestimation of Machine Learning Performance in Neuroimaging Studies of Depression}

\author[$\;\;$1,2]{\href{https://orcid.org/0000-0001-5164-8227}{Claas Flint \thanks{indicates that the authors contributed equally to the work and should be regarded as first authors.}}}
\author[$*$ 3,5]{\href{https://orcid.org/0000-0002-3353-8566}{Micah Cearns}}
\author[1]{\href{https://orcid.org/0000-0003-4749-3298}{Nils Opel}}
\author[1]{\href{https://orcid.org/0000-0002-7018-4525}{Ronny Redlich}}
\author[1]{\href{https://orcid.org/0000-0001-6587-2617}{David M. A. Mehler}}
\author[1]{\href{https://orcid.org/0000-0001-7459-6634}{Daniel Emden}}
\author[1]{\href{https://orcid.org/0000-0002-6241-1492}{Nils R. Winter}}
\author[1]{Ramona Leenings}
\author[4,8]{\href{https://orcid.org/0000-0001-6363-2759}{Simon B. Eickhoff}}
\author[6]{\href{https://orcid.org/0000-0002-2514-2625}{Tilo Kircher}}
\author[6]{\href{https://orcid.org/0000-0002-0564-2497}{Axel Krug}}
\author[6]{\href{https://orcid.org/0000-0002-0749-7473}{Igor Nenadic}}
\author[1]{\href{https://orcid.org/0000-0002-2445-9778}{Volker Arolt}}
\author[3]{Scott Clark}
\author[3,5,7]{\href{https://orcid.org/0000-0001-6548-426X}{Bernhard T. Baune}}
\author[2]{\href{https://orcid.org/0000-0001-7678-9528}{Xiaoyi Jiang}}
\author[$\;\;\ddagger$1]{\href{https://orcid.org/0000-0002-0623-3759}{Udo Dannlowski \thanks{indicates that the authors contributed equally to the work and should be regarded as senior authors.}}}
\author[$\;\dagger$1]{\href{https://orcid.org/0000-0002-8929-4134}{Tim Hahn}}

\affil[1]{Department of Psychiatry, University of Münster, Germany}
\affil[2]{Faculty of Mathematics and Computer Science, University of Münster, Germany}
\affil[3]{Discipline of Psychiatry, School of Medicine, University of Adelaide, Australia}
\affil[4]{Institute of Neuroscience and Medicine (INM-7) Research Center Jülich}
\affil[5]{Department of Psychiatry, Melbourne Medical School, The University of Melbourne, Parkville, Australia}
\affil[6]{Department of Psychiatry and Psychotherapy, University of Marburg, Germany}
\affil[7]{The Florey Institute of Neuroscience and Mental Health, The University of Melbourne, Parkville, Australia}
\affil[8]{Institute of Systems Neuroscience, Medical Faculty, Heinrich Heine University Düsseldorf, Düsseldorf, Germany}
\affil[$\ddagger$]{Corresponding author: Udo Dannlowski, Phone: +49-251-83-56610, Email: \href{mailto:dannlow@uni-muenster.de}{dannlow@uni-muenster.de}}

\renewcommand{\shorttitle}{Performance Misestimation in Neuroimaging Studies of Depression}
\hypersetup{
    pdftitle={Systematic Misestimation of Machine Learning Performance in Neuroimaging Studies of Depression},
    pdfauthor={
        Claas Flint,
        Micah Cearns,
        Nils Opel,
        Ronny Redlich,
        David M. A. Mehler,
        Daniel Emden,
        Nils R. Winter,
        Ramona Leenings,
        Simon B. Eickhoff,
        Tilo Kircher,
        Axel Krug,
        Igor Nenadic,
        Volker Arolt,
        Scott Clark,
        Bernhard T. Baune,
        Xiaoyi Jiang,
        Udo Dannlowski,
        Tim Hahn
    },
    pdfproducer={Claas Flint},
    pdfkeywords={machine learning, neuroimaging, major depression, misestimation, small sample size, clinical translation},
    colorlinks=false
}

\begin{document}
    \maketitle
    \begin{abstract}
        We currently observe a disconcerting phenomenon in machine learning studies in psychiatry: While we would expect larger samples to yield better results due to the availability of more data, larger machine learning studies consistently show much weaker performance than the numerous small-scale studies. Here, we systematically investigated this effect focusing on one of the most heavily studied questions in the field, namely the classification of patients suffering from \ac{mdd} and \ac{hc} based on neuroimaging data. Drawing upon structural \ac{mri} data from a balanced sample of $N$\,=\;\num{1868} \ac{mdd} patients and \ac{hc} from our recent international \ac{pac}, we first trained and tested a classification model on the full dataset which yielded an accuracy of \SI{61}{\percent}. Next, we mimicked the process by which researchers would draw samples of various sizes ($N$\,=\;\num{4} to $N$\,=\;\num{150}) from the population and showed a strong risk of misestimation. Specifically, for smallsample sizes ($N$\,=\;\num{20}), we observe accuracies of up to \SI{95}{\percent}. For medium sample sizes ($N$\,=\;\num{100}) accuracies up to \SI{75}{\percent} were found. Importantly, further investigation showed that sufficiently large test sets effectively protect against performance misestimation whereas larger datasets per se do not. While these results question the validity of a substantial part of the current literature, we outline the relatively low-cost remedy of larger test sets, which is readily available in most cases.

    \end{abstract}


    \pagebreak

    \section{Introduction}
    \acresetall
    In psychiatry, we are witnessing an explosion of interest in \ac{ml} and artificial intelligence for prediction and biomarker discovery, paralleling similar developments in personalized medicine~\cite{Darcy2016,Eyre2016,Gabrieli2015,Jordan2015}. In contrast to the majority of investigations employing classic group-level statistical inference, \ac{ml} approaches aim to build models which allow for individual (i.e.\ single subject) predictions, thus enabling direct assessment of individual differences and clinical utility~\cite{Hahn2017}. While this constitutes a major advancement for clinical translation, recent results of large-scale investigations have given rise to a fundamental concern in the field: Specifically, \ac{ml} studies including larger samples did not yield stronger performance, but consistently showed weaker results than studies drawing on small samples, calling into question the validity and generalizability of a large number of highly published proof-of-concept studies.

    The magnitude of this issue was impressively illustrated by the results of the \ac{pac} 2018 (Supplementary appendix~/ref{cha:pac-2018}) in which participants developed \ac{ml} models to classify \ac{hc} and patients diagnosed with \ac{mdd} based on structural \ac{mri} data from $N=\num{2240}$ participants. Despite the best efforts of $\sim\num{170}$ machine learners in \num{49} teams from around the world, performance ranged between \SI{60}{\percent} and \SI{65}{\percent} accuracy in a large, independent test set. This is in strong contrast to the numerous smaller studies showing accuracies of \SI{80}{\percent} or more~\cite{Johnston2015,Mwangi2012,Patel2015}.

    Further empirical studies focusing on other disorders support this observed effect of performance deterioration with increasing sample size: In a large-scale investigation, Neuhaus \& Popescu~\cite{Neuhaus2018} aggregated original studies across disorder categories, including schizophrenia(total observation $N=\num{5563}$), \ac{mdd} ($N=\num{2042}$), and \ac{adhd} ($N=\num{8084}$), finding an inverse relationship between sample size and balanced accuracy(schizophrenia, $r=\num{-.34}$; \ac{mdd}, $r=\num{-.32}$; and \ac{adhd}, $r=\num{-.43}$). Similar results were observed in a recent review of \num{200} neuroimaging classification studies of brain disorders, finding a general trend towards lower reported accuracy scores in larger samples~\cite{Arbabshirani2017}. Given that model performance would be expected to increase with more data, these results hint at a fundamental issue hampering current predictive biomarker studies in psychiatry.

    From a methodological point of view, it has been known since the early 90's that training samples should be large when there is a high number of features (i.e.\ measured variables) and a complex classification rule being fit to a dataset~\cite{Raudys1991}. Recent works have further reiterated this point~\cite{VanderPloeg2014}. Moreover, these effects may have been further amplified by certain cross-validation schemes. For example, Kambeitz et al.~\cite{Kambeitz2017} observed higher accuracy estimates in studies using hold-out cross-validation strategies compared to 10-fold and \ac{loocv}, whilst Varoquaux et al. observed that \ac{loocv} leads to unstable and biased estimates, concluding that repeated random splits should be preferred~\cite{Varoquaux2017}.

    Although these findings have sparked in-depth conceptual considerations~\cite{Hahn2019}, empirical investigations of this problem have been limited to specific cross-validation schemes and small test set sizes~\cite{Varoquaux2018}. Here, we aim to systematically investigate the effects of both train and test set sample sizes on \ac{ml} model performance in neuroimaging based \ac{mdd} classification. In addition, as it is possible that effects of systematic misestimation have arisen due to suboptimal pipeline configurations (i.e., the disproportionate use of \ac{loocv} and linear \acp{svm} on samples containing more predictors than observations), we also test a further \num{48} different pipeline configurations to quantify the influence of these additional factors. To demonstrate that this effect was not dependent on the data or any pipeline configurations used in our analyses, we repeat the process using a \emph{dummy classifier}. To quantify the magnitude of these effects in each configuration, we drew samples of various sizes from the \ac{pac} dataset - mimicking the process by which researchers would draw samples from the population of \ac{ml} studies reported in the literature. The resulting probability distributions are investigated.

    \section{Materials and methods}
    To investigate the effects of both train and test set sample sizes on \ac{ml} model performance in neuroimaging based \ac{mdd} classification, we repeatedly drew samples of different sizes from the \ac{pac} dataset to imitate the procedure reported in the literature. Subsequently, the resulting probability distributions are investigated.

    \subsection{Data description}
    The \ac{pac} dataset comprised anonymized, pre-processed \ac{mri} data of $N=\num{2240}$ individuals obtained from two large, independent, ongoing studies - the Münster Neuroimaging Cohort~\cite{Dannlowski2015,Dannlowski2015a} ($N=\num{724}$ \ac{mdd}; $N=\num{285}$ \ac{hc}) and the FOR2107-study~\cite{Kircher2018} ($N=\num{582}$ \ac{mdd}, $N=\num{649}$ \ac{hc}). Case/control status was diagnosed with the {SCID-IV}~\cite{Wittchen1997} interview employed by trained clinical raters in both studies. In both cohorts, exclusion criteria were any neurological or severe medical condition, \ac{mri} contraindications, substance-related disorders, Benzodiazepine treatment and head injuries. For healthy controls, any current or previous psychiatric disorder or use of any psychotropic substances. The Münster Neuroimaging Cohort was scanned at one single \ac{mri}-scanner with the same sequence, while the FOR2107-study was scanned at two independent sites~\cite{Vogelbacher2018}, yielding \num{3} different scanner types/sequences. The structural T1-weighted \ac{mri} scans were pre-processed with the CAT12 toolbox (\url{http://www.neuro.uni-jena.de/cat}, r1184) using default parameters to obtain modulated, normalized grey matter segments (resolution $\num{1.5} \times \num{1.5} \times \SI{1.5}{\cubic\milli\meter}$) which were used for the present analysis. Furthermore, age, gender, scanner type, and \ac{tiv} were provided.

    The FOR2107 cohort project was approved by the Ethics Committees of the Medical Faculties, University of Marburg and University of Münster.

    \subsection{Machine learning pipeline}
    To ensure the unbiased approximation of the model's performance in previously unseen patients (i.e., model generalization), we trained and tested all models in a pipeline to prevent information leaking between patients used for training and validation. To avoid a confounding effect due to an imbalance in the sample, we used random under-sampling in a first step to obtain a balanced sample of \num{934} \ac{mdd} cases and \num{934} \ac{hc} (see Supplementary appendix~\ref{cha:summary-statistics} for summary statistics). To reduce the computational effort, all images have been scaled down to a voxel size of $\num{3} \times \num{3} \times \SI{3}{\cubic\milli\meter}$. Following, every image was converted to a vector, where every voxel served as a feature. After background elimination (features with no variance), \num{58255} features remained. For standardization, the features were scaled to have zero mean and unit variance. Finally, a linear \ac{svm} with default parameters (Scikit-learn~\cite{scikit-learn}, v0.20) was trained and model performance was calculated and analysed in the subsequent analyses. The effect of varying scanner-distributions across samples was found to be negligible (see Supplementary appendix~\ref{cha:influence-of-scanner-distribution}).

    \subsection{Overall sample size effects}
    To first examine the effects of overall-set size, we randomly sampled the \ac{pac} dataset in steps of \num{1} from $N=4$ to $N=150$. For each $N$ value we created \num{1000} balanced samples. This yielded \num{147000} random samples with equal numbers of \ac{mdd} patients and \ac{hc}. Applying the \ac{svm} pipeline described above to each of these samples independently allowed us to obtain a distribution of accuracy scores for each $N$ value, respectively. On each sample one \ac{svm} was trained with default parameters using \ac{loocv}. Thus, we trained a total of \num{11315000} \acp{svm}. Since the computational effort increases quadratically with increasing sample size, and in addition, the most recent neuroimaging \ac{ml} studies using \ac{loocv} rarely exceed $N=150$, we decided to stop at this value. Following, we evaluated the distribution of accuracy scores estimated for each sample size ($N=4$ to $150$).

    \subsection{Training set size effects}
    Second, to examine the effects of training set size, we varied the size of each training set and then tested performance on a fixed hold-out set. Specifically, we randomly sampled the \ac{pac} dataset in steps of $1$ from $N=4$ to $150$ as was done in the previous analysis. For each $N$ value, we created \num{1000} balanced samples and used each of them to train different models. To then test the effects of training set size on test set performance, we tested each trained model on a balanced test set of $N=300$. From the resulting distributions, we quantified the probability of overestimating accuracy as a function of training set size.

    \subsection{Test set size effects}
\begin{figure}
        \includegraphics[width=\textwidth]{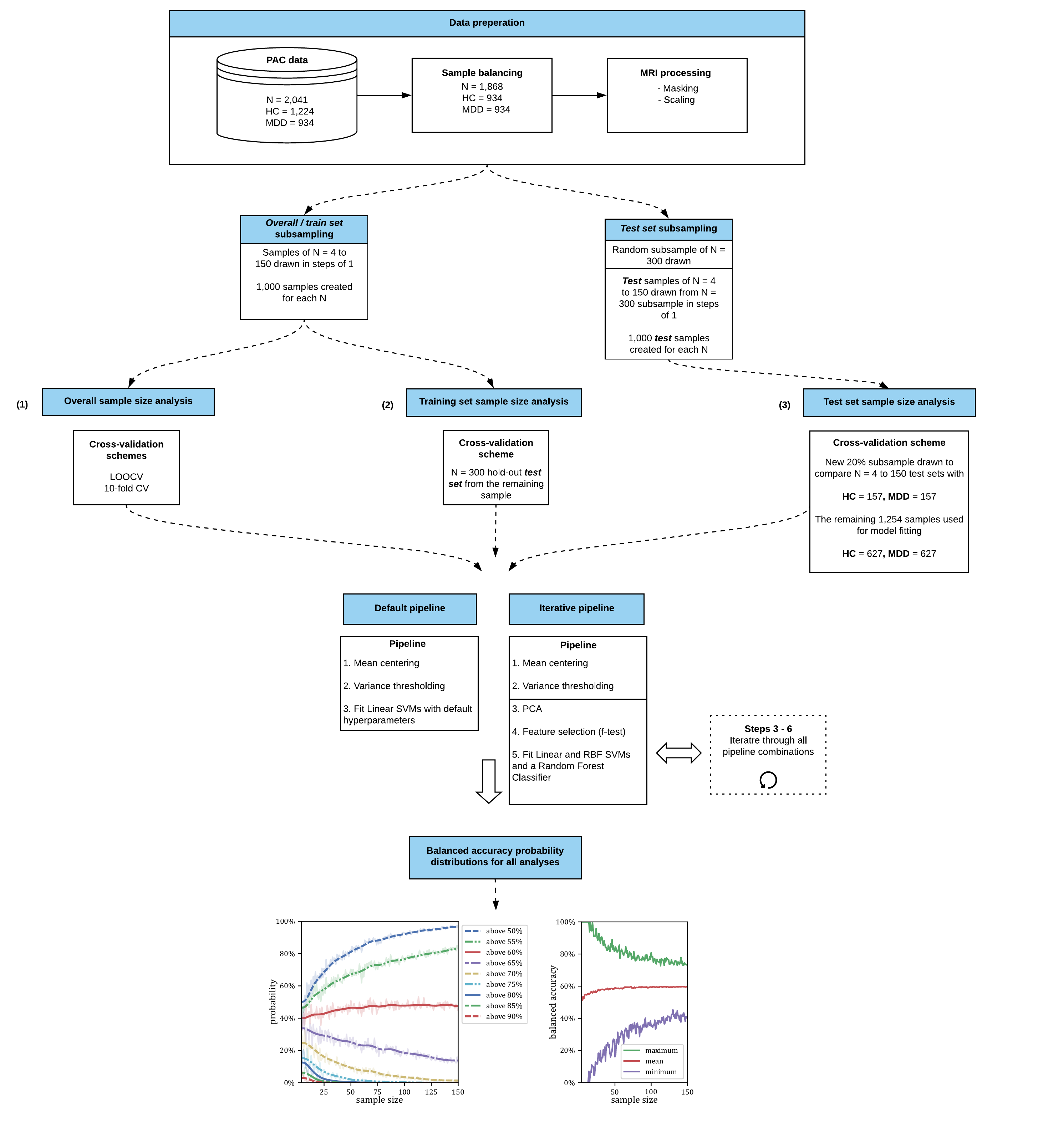}
        \caption{Workflow to investigate the correlations between sample size and misestimation. First the effect of misestimation is investigated over the whole classification process ((1) Overall sample size analysis). Following the process of training and testing is evaluated separately ((2) Training set sample size analysis, (3) Test set sample size analysis)}
        \label{fig:workflow}
    \end{figure}

    To assess the effects of test set size on classification accuracy, we randomly sampled a balanced group of \num{300} subjects from our full sample of \num{934} \ac{mdd} cases and \num{934} \acp{hc}. From this sample, we randomly sub-sampled test sets of $N=\num{4}$ to \num{150} in steps of \num{1}. For each value of $N$, we took \num{1000} samples, resulting in the creation of \num{147000} random test samples. From the remaining \num{784} \ac{mdd} cases and \num{784} \acp{hc}, we took out another \SI{20}{\percent} sample ($\text{\ac{mdd}}=157$ and $\text{\ac{hc}}=157$). The prediction of this sample provides a reliable basis for an overall performance estimation and allows for a comparison with the results obtained from smaller samples. We then trained a single \ac{svm} on the remaining sample ($\text{\ac{mdd}}=627$, $\text{\ac{hc}}=627$). The trained \ac{svm} was then tested on each test set sample ($N=\num{4}$ to \num{150}). From the resulting test set accuracy distributions, we derived the probability of obtaining accuracy scores between \SI{50}{\percent} and \SI{90}{\percent} accuracy by chance. See Figure~\ref{fig:workflow} for an overview of all analyses.

    \subsection{Generalizability of statistical effect}

    \subsubsection{Alternative pipeline configurations}
    As we have attempted to hold other components of the modelling process constant so any observed effects of systematic misestimation can be attributed to sample size alone, it is also possible that our results may be dependent on the basic configuration of our \ac{ml} pipeline (e.g., the use of a linear \ac{svm}, default hyperparameters, and \ac{loocv}). Therefore, we trained a further \num{48} pipeline configurations, including the use of both linear and radial basis function \acp{svm} and a Random Forest classifier, all of which have demonstrated their efficacy in neuroimaging classification studies~\cite{Kambeitz2017}. Within these configurations, we conducted dimensionality reduction using \ac{pca} as well as f-test based feature selection. This allowed us to assess whether our findings were being confounded by the large number of predictors used in the main analysis. For further information see Supplement Append~x~\ref{cha:alternative-machine-configurations} and for all results see supplementary results~\ref{cha:alternative-machine-configuration-results}, supplementary figures~\ref{fig:no_PCA_no_selection_RandomForest} -~\ref{fig:PCA_all_components_1000_best_selected_LinearSVC} and supplementary tables~\ref{tab:no_PCA_no_selection_RandomForest}~-~\ref{tab:PCA_all_components_1000_best_selected_LinearSVC}.

    \subsubsection{Dummy classifier}
In order to show that the observed effects were not specific to the \ac{pac} dataset or any of the alternative pipeline configurations in these analyses, we repeated the procedures described above with a dummy classifier. Our dummy classifier assumed a prior probability for \ac{mdd} vs \ac{hc} classification based on the percentage proportion of each class in the training data (prevalence). As the dataset was balanced with random under-sampling, the prior and subsequent ground truth of the model was equal to exactly \SI{50}{\percent}. This approach allowed us to compare our distribution of dummy performance estimates derived from our subsample analysis to this ground truth value. Importantly, this approach allows for the quantification of accuracy misestimation as a function of sample size completely independent of any unique characteristics that may be specific to the \ac{pac} dataset or our pipeline configurations. Therefore, we can then be sure that any subsequent changes in classifier performance are attributable to sample size alone.

    \section{Results}

    \subsection{Overall sample size effects}
\begin{figure}
        \captionsetup[subfigure]{justification=justified,singlelinecheck=false}
        \begin{subfigure}[t]{0.61\textwidth}
            \input{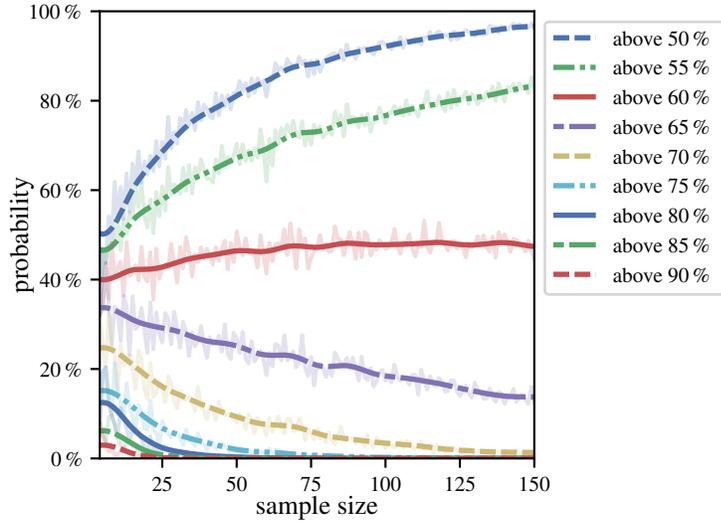}
            \caption{Probabilities for linear \acp{svm} to yield an accuracy exceeding a certain threshold as a function of sample size employing \ac{loocv}. }
        \end{subfigure}
        \hspace{3.0mm}
        \begin{subfigure}[t]{0.34\textwidth}
            \input{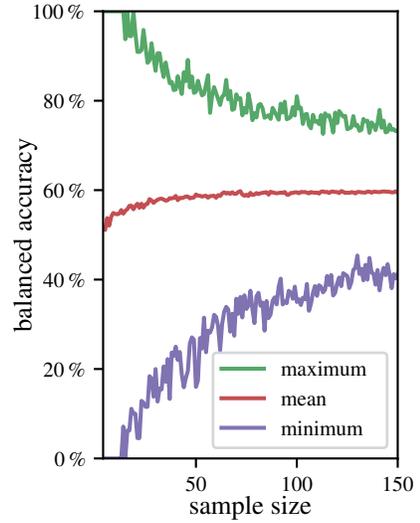}
            \caption{Minimum, maximum and mean results for the linear \acp{svm} as a function of sample size employing \ac{loocv}.}
        \end{subfigure}

        \vspace{3.0mm}

        \begin{subfigure}[t]{0.61\textwidth}
            \input{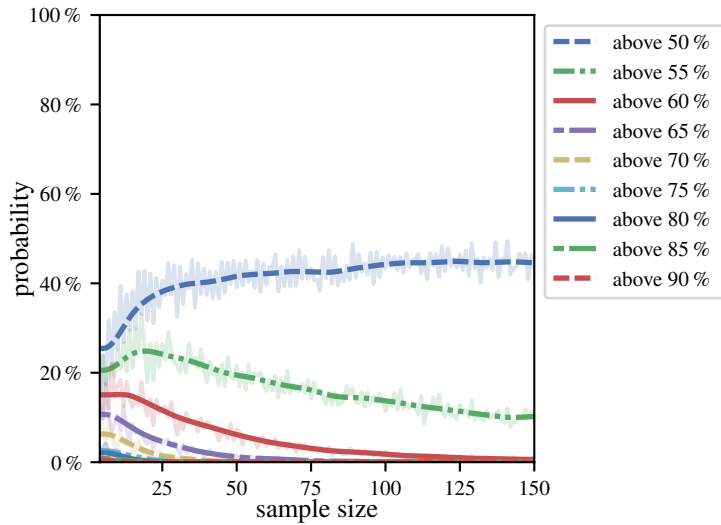}
            \caption{Probabilities for dummy classifiers’s to get an accuracy above a certain chance level related to the size of the used sample.}
        \end{subfigure}
        \hspace{3.0mm}
        \begin{subfigure}[t]{0.34\textwidth}
            \input{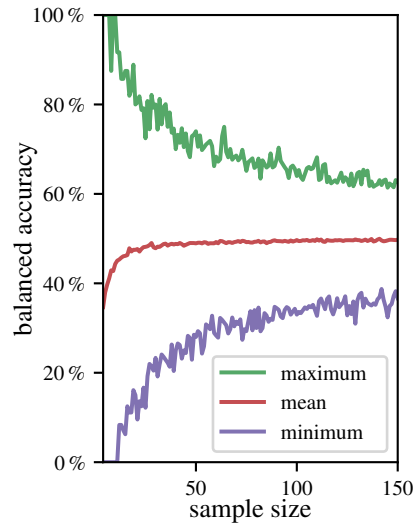}
            \caption{Minimum, maximum and mean results for the dummy classifiers related to the size of the used sample size for training and testing.}
        \end{subfigure}
        \caption{Effects of varying overall sample sizes employing \ac{loocv}.}
        \label{fig:overall_sample_size_effects}
    \end{figure}

    In the overall sample size analysis using \ac{loocv}, we were able to show that the risk of overestimating the classifier performance increases with decreasing overall sample size (Figure~\ref{fig:overall_sample_size_effects}a, Supplements Table~\ref{tab:overall_linear_svm_loocv}a). Specifically, accuracies of \SI{70}{\percent} or higher are observed with a probability of \SI{13}{\percent} on sample sizes of $N=20$ whereas this probability is reduced to \SI{2}{\percent} for sample sizes of $N=100$. In addition, sample size has a profound impact on the variability of accuracy estimates: For samples of size $N=20$, accuracies ranged from \SI{10}{\percent} to \SI{95}{\percent} ($\text{standard deviation}=\SI{15}{\percent}$) while for samples of $N=100$, accuracies ranged between \SI{35}{\percent} and \SI{81}{\percent} ($\text{standard deviation}=\SI{6}{\percent}$) (Figure~\ref{fig:overall_sample_size_effects}b, Supplements Table~\ref{tab:overall_linear_svm_loocv}b). Note that this effect is symmetrical and also applies to the underestimation of performance (Figure~\ref{fig:overall_sample_size_effects}b). Additionally, the results of the dummy classifier (Figure~\ref{fig:overall_sample_size_effects}c, d, Supplements Table~\ref{tab:overall_dummy_classifier_loocv}) show that the observed overestimation effect is a general effect of sample size as previously pointed out by Varoquaux~\cite{Varoquaux2018}. As the regularization of the \ac{svm} is sensitive to the total number of outliers, which may increase in parallel with sample size, we conducted an additional analysis with adjusted $C$ parameters, with the observed effect remaining constant across these analyses (see Supplements Figure~\ref{fig:adj_c_overall} and~\ref{fig:adj_c_test}).

    \subsection{Training set size effects}
\begin{figure}
        \captionsetup[subfigure]{justification=justified,singlelinecheck=false}
        \begin{subfigure}[t]{0.61\textwidth}
            \input{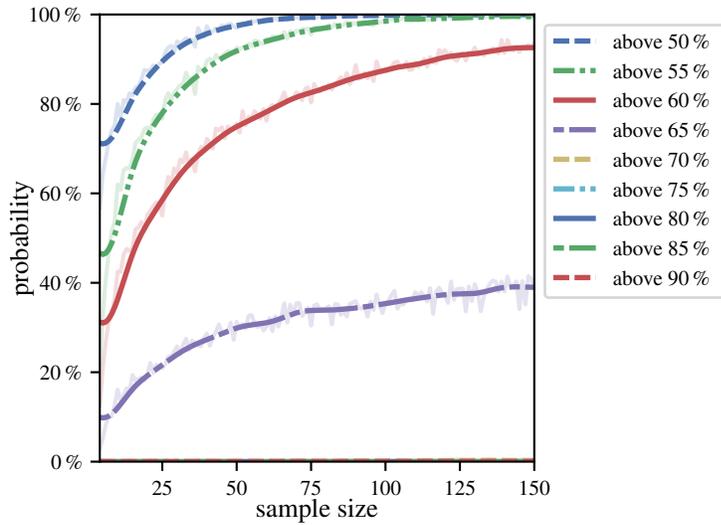}
            \caption{Probabilities for linear \acp{svm} to yield an accuracy exceeding a certain threshold as a function of training sample size. }
        \end{subfigure}
        \hspace{3.0mm}
        \begin{subfigure}[t]{0.34\textwidth}
            \input{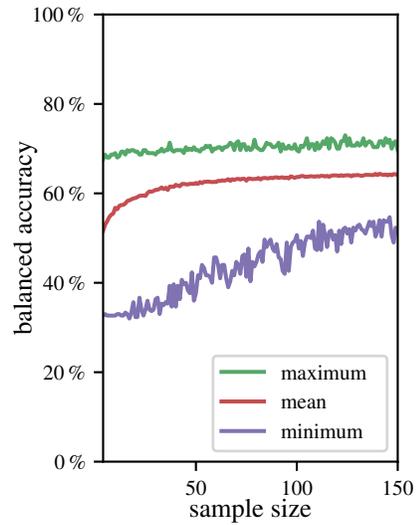}
            \caption{Minimum, maximum and mean results for the linear \acp{svm} as a function of training sample size. }
        \end{subfigure}

        \vspace{3.0mm}

        \begin{subfigure}[t]{0.61\textwidth}
            \input{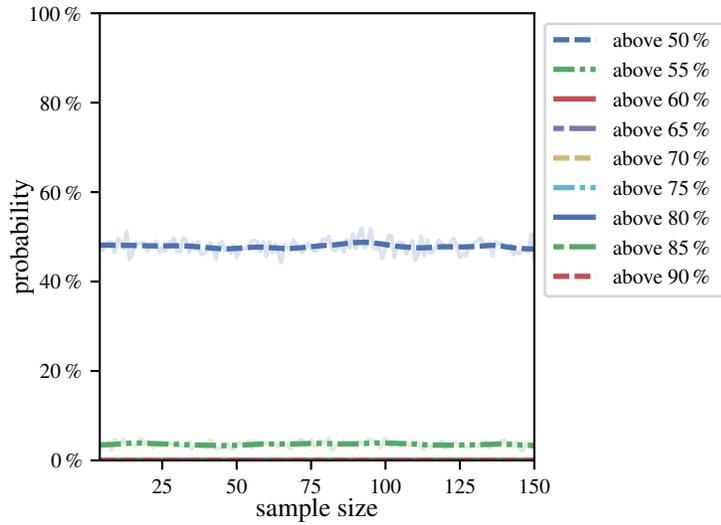}
            \caption{Probabilities for the dummy classifier to get an accuracy above a certain chance level related to the size of the training set size.}
        \end{subfigure}
        \hspace{3.0mm}
        \begin{subfigure}[t]{0.34\textwidth}
            \input{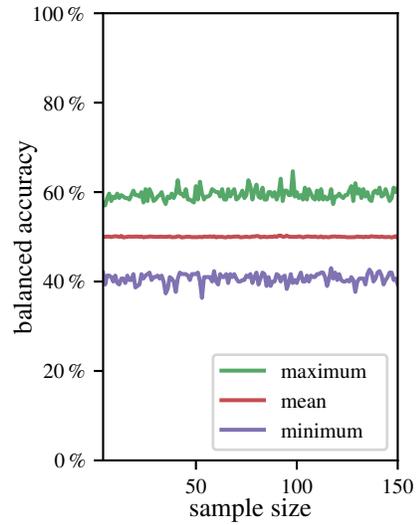}
            \caption{Minimum, maximum and mean results for the dummy classifier related to the size of the used sample size for training.}
        \end{subfigure}
        \caption{Results as a function of training set sizes with a fixed test set size of $N = \num{300}$.}
        \label{fig:train_sample_size_effects}
    \end{figure}

    When examining the effects of training set size ($N=\num{4}$ to \num{150}) using a large test set for evaluation ($N=\num{300}$), we did not observe any systematic misestimation (Figure~\ref{fig:train_sample_size_effects}a, Supplements Table~\ref{tab:train_set_linear_svm}a). In fact, models trained on virtually any training set size from $N=\num{4}$ to \num{150} were sufficient to obtain maximum model performance. However, increasing training sample size decreased the probability of obtaining very low performance estimate. For training sets of size $N=\num{20}$, accuracies ranged from \SI{32}{\percent} to \SI{69}{\percent} ($\text{standard deviation}=\SI{7.1}{\percent}$) while for training sets of $N=100$, accuracies ranged between \SI{51}{\percent} and \SI{70}{\percent} ($\text{standard deviation}=\SI{3.0}{\percent}$) (Figure~\ref{fig:train_sample_size_effects}b, Supplements Table~\ref{tab:train_set_linear_svm}b). In accordance with the overall sample size analysis, the results of the dummy classifier (Figure~\ref{fig:train_sample_size_effects}c, d, Supplements Table~\ref{tab:train_dummy_classifier}) showed that this observed effect was general in nature.

    \subsection{Test set size effects}
\begin{figure}
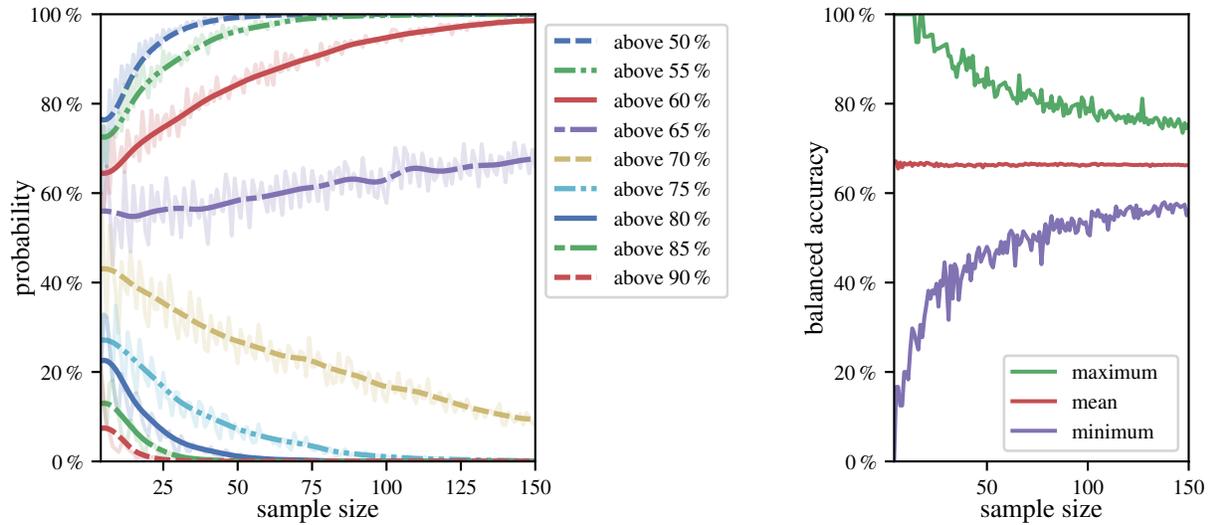
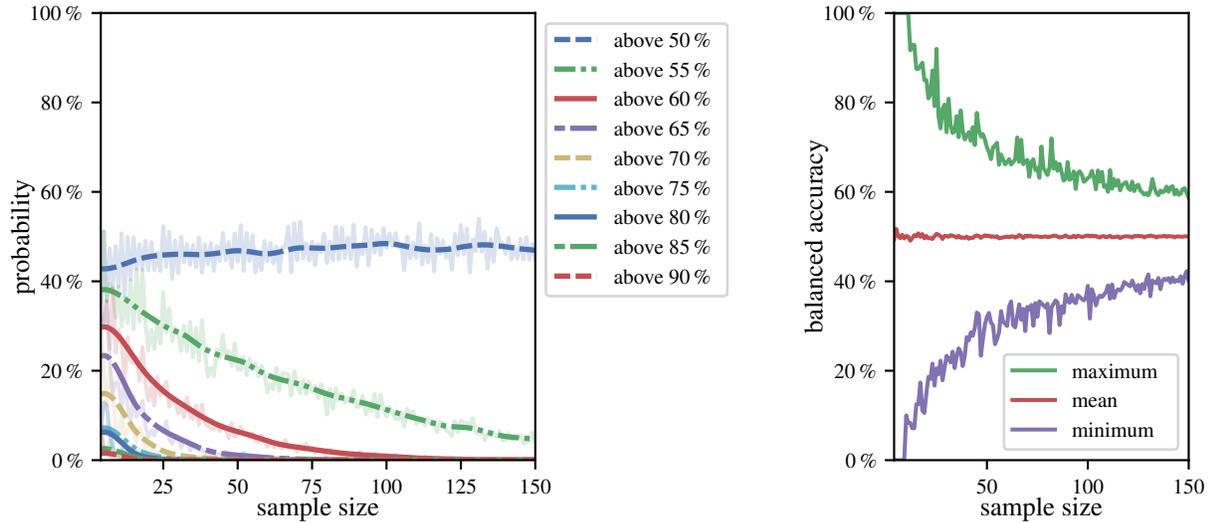

        \captionsetup[subfigure]{justification=justified,singlelinecheck=false}
        \begin{subfigure}[t]{0.61\textwidth}
            \input{images/test_set_size_svm_chances.pgf}
            \caption{Probabilities for linear \acp{svm} to yield an accuracy exceeding a certain threshold as a function of test sample size.}
        \end{subfigure}
        \hspace{3.0mm}
        \begin{subfigure}[t]{0.34\textwidth}
            \input{images/test_set_size_svm_stats.pgf}
            \caption{Minimum, maximum and mean results for the linear \acp{svm} as a function of test sample size.}
        \end{subfigure}

        \vspace{3.0mm}

        \begin{subfigure}[t]{0.61\textwidth}
            \input{images/test_set_size_dummy_chances.pgf}
            \caption{Probabilities for the dummy classifier to get an accuracy above a certain chance level related to the size of the test set size.}
        \end{subfigure}
        \hspace{3.0mm}
        \begin{subfigure}[t]{0.34\textwidth}
            \input{images/test_set_size_dummy_stats.pgf}
            \caption{Minimum, maximum and mean results for the dummy classifier related to the size of the used sample size for testing.}
        \end{subfigure}
        \caption{Results as a function of variable test set sizes with and a fixed classifier.}
        \label{fig:test_sample_size_effects}
    \end{figure}

    For our analysis varying test set size, we found a similar pattern of systematic misestimation as that in our first overall sample size analysis using \ac{loocv}. With a sample size of $N=\num{20}$, we obtain results of \SI{70}{\percent} accuracy or higher with a probability of \SI{30}{\percent}, whilst the mean accuracy on the full dataset ($N=268$) was only \SI{61}{\percent}. This probability dropped to \SI{13}{\percent} when the test sample size was $N=100$. For test sets of size $N=20$, accuracies ranged from \SI{35}{\percent} to \SI{95}{\percent} ($\text{standard deviation}=\SI{10}{\percent}$) whilst for test sets of size $N=100$, accuracies ranged from \SI{51}{\percent} and \SI{79}{\percent} ($\text{standard deviation}=\SI{4}{\percent}$) (Figure~\ref{fig:test_sample_size_effects}b, Supplements Table~\ref{tab:test_set_dummy_classifier}b). Running the analysis again using our dummy classifier, we were able to show that the general pattern of systematic overestimation was independent of the specific dataset used (Figure~\ref{fig:test_sample_size_effects}c, d, Supplements Table~\ref{tab:test_set_dummy_classifier}). To show the independent and generalizable character of the observed effect, we repeated the analysis on the \num{48} unique pipeline configurations discussed above (see Supplementary appendix~\ref{cha:alternative-machine-configurations}). Specifically, the results are comparable to the original used configuration, i.e.\ an \ac{svm} with a linear kernel and no preprocessing. Finally, our analysis of scanner sites revealed no effects on model performance (see Supplementary appendix~\ref{cha:adjustment-svm-regularization}).

    \section{Discussion}
Sparked by the observation that machine learning studies drawing on larger neuroimaging samples consistently showed weaker results than studies drawing on smaller ones, we drew samples of various sizes from the \ac{pac} dataset, thereby mimicking the process by which researchers would draw samples from the population of \ac{ml} studies reported in the literature. When applying a linear \ac{svm} with \ac{loocv}, as is the most common approach in the neuroimaging literature~\cite{Arbabshirani2017}, we observed a higher risk of misestimations, which may lead to artificially high performance estimates in smaller samples. Importantly, our analyses revealed that this is primarily due to a small test set, not training set size. Generally, this shows that a small test sample size may explain many of the highly optimistic results published in recent years. When considering the well-established effect of publication bias, even if underestimated results are equally as likely, they will have a significantly lower chance of being published.

    Our results are the first to disentangle the effects of training and test set size effects which are typically inseparable in common cross-validation frameworks such as \ac{loocv}. This delineation of effects enabled two important insights for biomarker discovery and outcome prediction. First, researchers need to validate their models on large, independent test sets. Our results indicate that in the \ac{pac} dataset, a test set size of $N=\num{100}$ was already sufficient to lower the probability of obtaining artificially good performance (i.e., \SI{70}{\percent} or higher) to \SI{13}{\percent}. With a median $N$ of less than \num{100} in many published studies~\cite{Arbabshirani2017}, this may seem challenging. However, online infrastructure for independent machine learning model evaluation is available (e.g., \url{www.photon-ai.com/repo}). If researchers open-source their models, anyone - independent of technical knowledge or machine learning expertise - can evaluate them. This way, large independent test datasets can be obtained in a short time without the need for data sharing. This is not restricted to neuroimaging data, but any machine learning model. In addition, efforts from consortia will also help to mitigate this problem and should be considered by machine learning practitioners.

Second, the size of the training set alone cannot serve as an indicator of later model performance. Larger training sets are more likely to generalize to new data and broaden model scope (i.e., about which groups within a population a given model can make reasonable predictions), however, in the current analysis, the linear rule learned by an \ac{svm} on high-dimensional neuroimaging data could be approximated with only a handful of samples. From a training set size of \num{30} onward, we no longer observed any increase in model performance. This somewhat counterintuitive effect arises whenever a simple rule is approximated. For higher complexity models (i.e., models capable of learning more complex rules), we, of course, expect performance increases as training sample size increases. However, considering the results of the \ac{pac} (Supplementary appendix~\ref{cha:pac-2018}), high complexity models such as Deep Learning approaches did not yield higher performance when trained with $\sim\num{1000}$ samples. Thus, we conclude that simple models are competitive for sample sizes of up to $N=\num{1000}$ for this particular classification problem. Whether more complex rules can be discovered beyond this point or whether other fundamental issues hamper biomarker discovery in psychiatry (cf. e.g.\ biotype identification~\cite{Kircher2018} and normative modelling approaches~\cite{Marquand2016}) remains an open question.

An intuitive criticism of our main analyses would be that it has merely replicated methods similar to those of previous low-quality works (for example, studies using only linear \acp{svm} with default parameters, tested in \ac{loocv} schemes, on samples with many more predictors than observations). Whilst this pipeline configuration was used in the current analysis to a) hold constant properties, that if varied, may have been indistinguishably responsible for changes in accuracy misestimation, and b) replicate the most commonly used \ac{ml} pipeline configuration in our field, it was important to conduct complementary analyses to rule out these confounds. Therefore, we tested a further \num{48} \ac{ml} pipeline configurations (see Supplementary appendix~\ref{cha:alternative-machine-configurations}) using both linear (a linear \ac{svm}) and non-linear (a \ac{rbf} \ac{svm} and a Random Forest) classifiers. In addition, we conducted \ac{pca} based dimensionality reduction as well as f-test based feature selection within these classifiers to delineate whether our findings were confounded by the large size of the predictor space relative to our number of observations. Importantly, all pipeline configurations demonstrated the same pattern of systematic misestimation as that in our main analyses. The second potential criticism of the current work is that these findings may merely be modality-specific, limiting the generalizability of these findings across domains. However, the use of a dummy classifier that completely ignored the input predictor space (the voxels), and instead, classified samples based only on their prevalence in training ($\text{\ac{mdd}}=\SI{50}{\percent}$, $\text{\ac{hc}}=\SI{50}{\percent}$), showed the same pattern of sample size based systematic misestimation across all pipeline configurations, thus, demonstrating a generalizable statistical effect regardless of the data modality used.

    Given the profound effect of the test set size on systematic misestimation, it is important to consider why an effect of overestimation may arise. Previous work by Schnack \& Kahn~\cite{Schnack2016} suggests that patient characteristics in smaller samples tend to be more homogeneous. In the case of small $N$, participants may be more likely to be recruited from the same data collection site and of a similar age (for example, in the case of a university recruited convenience sample). In addition, stringent recruitment criterion may be easily met, resulting in a well-defined phenotype that is not truly representative of the parent population of interest. Whilst this explanation makes sense for samples collected in this manner, it fails to explain why we observed this phenomenon in our random sampling procedure, and more importantly, with our dummy classifiers that paid no attention to participants, their characteristics, or the inputted predictor variables. This observation suggests a mechanism for systematic misestimation that is not just sample/patient-specific or contingent on sample homogeneity, but instead, inherent in the natural variation that arises in small test samples. Indeed, this effect is known as sampling error, and as demonstrated by Combrisson et al.~\cite{Combrisson2015} can lead to an effect whereby we exceed a machine learning model's chance level, purely by chance.

    In addition to sample size, other issues such as data leakage~\cite{Kambeitz2017}, are likely contributing to the systematic overestimation seen in the literature. Dedicated cross-platform software to help avoid data leakage is freely available (e.g.\ PHOTON, \url{www.photon-ai.com} or Scikit-learn~\cite{scikit-learn}). Finally, code should be made available on request or provided in an online repository (e.g., GitHub or GitLab) upon submission for review. In addition, a more elaborate evaluation framework including the analysis of model scope assessment as well as incremental utility and risk analysis is needed to move the field beyond proof-of-concept studies. The success of current translational efforts in neuroimaging and psychiatry will crucially depend on the timely adoption of guidelines and rules for medical machine learning models (for an in-depth introduction, see~\cite{Hahn2019}).

In summary, our results indicate that - while many of the most highly published results might strongly overestimate true performance - evaluation on large test sets constitutes a straightforward remedy. Given that simple, low-complexity models such as linear \acp{svm} did not gain from larger training set size, researchers should not discard their models due to low training $N$ but seek evaluation on a large test set for any model showing good performance.

    \section*{Acknowledgements}
    \subsection*{FOR2107}
    This work was funded by the German Research Foundation (DFG, grant FOR2107 DA1151/5-1 and DA1151/5-2 to UD; SFB-TRR58, Projects C09 and Z02 to UD) and the Interdisciplinary Center for Clinical Research (IZKF) of the medical faculty of Münster (grant Dan3/012/17 to UD). TH was supported by the German Research Foundation (DFG grants HA7070/2-2, HA7070/3, HA7070/4).

    This work is part of the German multicenter consortium "Neurobiology of Affective Disorders. A translational perspective on brain structure and function", funded by the German Research Foundation (Deutsche Forschungsgemeinschaft DFG; Forschungsgruppe/Research Unit FOR2107).

    \paragraph{Principal investigators (PIs) with respective areas of responsibility in the FOR2107 consortium are:}
    Work Package WP1, FOR2107/MACS cohort and brain imaging: Tilo Kircher (speaker FOR2107; DFG grant numbers KI 588/14-1, KI 588/14-2), Udo Dannlowski (co-speaker FOR2107; DA 1151/5-1, DA 1151/5-2), Axel Krug (KR 3822/5-1, KR 3822/7-2), Igor Nenadic (NE 2254/1-2), Carsten Konrad (KO 4291/3-1). WP2, animal phenotyping: Markus Wöhr (WO 1732/4-1, WO 1732/4-2), Rainer Schwarting (SCHW 559/14-1, SCHW 559/14-2). WP3, miRNA: Gerhard Schratt (SCHR 1136/3-1, 1136/3-2). WP4, immunology, mitochondriae: Judith Alferink (AL 1145/5-2), Carsten Culmsee (CU 43/9-1, CU 43/9-2), Holger Garn (GA 545/5-1, GA 545/7-2). WP5, genetics: Marcella Rietschel (RI 908/11-1, RI 908/11-2), Markus Nöthen (NO 246/10-1, NO 246/10-2), Stephanie Witt (WI 3439/3-1, WI 3439/3-2). WP6, multi method data analytics: Andreas Jansen (JA 1890/7-1, JA 1890/7-2), Tim Hahn (HA 7070/2-2), Bertram Müller-Myhsok (MU1315/8-2), Astrid Dempfle (DE 1614/3-1, DE 1614/3-2). CP1, biobank: Petra Pfefferle (PF 784/1-1, PF 784/1-2), Harald Renz (RE 737/20-1, 737/20-2). CP2, administration. Tilo Kircher (KI 588/15-1, KI 588/17-1), Udo Dannlowski (DA 1151/6-1), Carsten Konrad (KO 4291/4-1).

    \paragraph{Data access and responsibility:}
    All PIs take responsibility for the integrity of the respective study data and their components. All authors and coauthors had full access to all study data.

    \paragraph{Acknowledgements and members by Work Package (WP):}

    \subparagraph{WP1:}
    Henrike Bröhl, Katharina Brosch, Bruno Dietsche, Rozbeh Elahi, Jennifer Engelen, Sabine Fischer, Jessica Heinen, Svenja Klingel, Felicitas Meier, Tina Meller, Torsten Sauder, Simon Schmitt, Frederike Stein, Annette Tittmar, Dilara Yüksel (Dept. of Psychiatry, Marburg University). Mechthild Wallnig, Rita Werner (Core-Facility Brainimaging, Marburg University). Carmen Schade-Brittinger, Maik Hahmann (Coordinating Centre for Clinical Trials, Marburg). Michael Putzke (Psychiatric Hospital, Friedberg). Rolf Speier, Lutz Lenhard (Psychiatric Hospital, Haina). Birgit Köhnlein (Psychiatric Practice, Marburg). Peter Wulf, Jürgen Kleebach, Achim Becker (Psychiatric Hospital Hephata, Schwalmstadt-Treysa). Ruth Bär (Care facility Bischoff, Neunkirchen). Matthias Müller, Michael Franz, Siegfried Scharmann, Anja Haag, Kristina Spenner, Ulrich Ohlenschläger (Psychiatric Hospital Vitos, Marburg). Matthias Müller, Michael Franz, Bernd Kundermann (Psychiatric Hospital Vitos, Gießen). Christian Bürger, Fanni Dzvonyar, Verena Enneking, Stella Fingas, Janik Goltermann, Hannah Lemke, Susanne Meinert, Jonathan Repple, Kordula Vorspohl, Bettina Walden, Dario Zaremba (Dept. of Psychiatry, University of Münster). Harald Kugel, Jochen Bauer, Walter Heindel, Birgit Vahrenkamp (Dept. of Clinical Radiology, University of Münster). Gereon Heuft, Gudrun Schneider (Dept. of Psychosomatics and Psychotherapy, University of Münster). Thomas Reker (LWL-Hospital Münster). Gisela Bartling (IPP Münster). Ulrike Buhlmann (Dept. of Clinical Psychology, University of Münster).

    \subparagraph{WP2:} Marco Bartz, Miriam Becker, Christine Blöcher, Annuska Berz, Moria Braun, Ingmar Conell, Debora dalla Vecchia, Darius Dietrich, Ezgi Esen, Sophia Estel, Jens Hensen, Ruhkshona Kayumova, Theresa Kisko, Rebekka Obermeier, Anika Pützer, Nivethini Sangarapillai, Özge Sungur, Clara Raithel, Tobias Redecker, Vanessa Sandermann, Finnja Schramm, Linda Tempel, Natalie Vermehren, Jakob Vörckel, Stephan Weingarten, Maria Willadsen, Cüneyt Yildiz (Faculty of Psychology, Marburg University).

    \subparagraph{WP4:} Jana Freff, Silke Jörgens, Kathrin Schwarte (Dept. of Psychiatry, University of Münster). Susanne Michels, Goutham Ganjam, Katharina Elsässer (Faculty of Pharmacy, Marburg University). Felix Ruben Picard, Nicole Löwer, Thomas Ruppersberg (Institute of Laboratory Medicine and Pathobiochemistry, Marburg University).

    \subparagraph{WP5:} Helene Dukal, Christine Hohmeyer, Lennard Stütz, Viola Schwerdt, Fabian Streit, Josef Frank, Lea Sirignano (Dept. of Genetic Epidemiology, Central Institute of Mental Health, Medical Faculty Mannheim, Heidelberg University).

    \subparagraph{WP6:} Anastasia Benedyk, Miriam Bopp, Roman Keßler, Maximilian Lückel, Verena Schuster, Christoph Vogelbacher (Dept. of Psychiatry, Marburg University). Jens Sommer, Olaf Steinsträter (Core-Facility Brainimaging, Marburg University). Thomas W.D. Möbius (Institute of Medical Informatics and Statistics, Kiel University).

    \subparagraph{CP1:} Julian Glandorf, Fabian Kormann, Arif Alkan, Fatana Wedi, Lea Henning, Alena Renker, Karina Schneider, Elisabeth Folwarczny, Dana Stenzel, Kai Wenk, Felix Picard, Alexandra Fischer, Sandra Blumenau, Beate Kleb, Doris Finholdt, Elisabeth Kinder, Tamara Wüst, Elvira Przypadlo, Corinna Brehm (Comprehensive Biomaterial Bank Marburg, Marburg University).

    The FOR2107 cohort project (WP1) was approved by the Ethics Committees of the Medical Faculties, University of Marburg (AZ: 07/14) and University of Münster (AZ: 2014--422-b-S).

    \section*{Conflict of Interest}
    Biomedical financial interests or potential conflicts of interest: Tilo Kircher received unrestricted educational grants from Servier, Janssen, Recordati, Aristo, Otsuka, neuraxpharm.

    The other authors (Claas Flint, Micah Cearns, Nils Opel, Ronny Redlich, David M. A. Mehler, Daniel Emden, Nils R. Winter, Ramona Leenings, Simon B. Eickhoff, Axel Krug, Igor Nenadic, Volker Arolt, Scott Clark, Bernhard T. Baune, Xiaoyi Jiang, Udo Dannlowski, Tim Hahn) declare no conflicts of interest.

    \bibliographystyle{unsrt}
    \bibliography{references}

    %%%%%%%%%% Merge with supplemental materials %%%%%%%%%%
    \newpage
    \appendix

    \lhead{\scshape \footnotesize Supplements}

    \renewcommand{\shorttitle}{Performance Misestimation in Neuroimaging Studies of Depression}
    \begin{center}
        \LARGE{\textsc{Supplemental Materials}} \\ \Large{Systematic Misestimation of Machine Learning Performance \\ in Neuroimaging Studies of Depression}
    \end{center}

    \setcounter{equation}{1}
    \setcounter{figure}{0}
    \setcounter{table}{0}
    \setcounter{page}{1}
    \renewcommand\thefigure{\thesection.\arabic{figure}}
    \renewcommand\thetable{\thesection.\arabic{table}}

    %! language = Latex
\section{Summary statistics for the final study sample}
\label{cha:summary-statistics}
\begin{table}[!htp]
    \begin{center}
        \begin{subtable}[c]{0.6\textwidth}
            \begin{center}
                \begin{tabular}{rc|cc}
                    &        & {\textbf{\acs{mdd}}} & {\textbf{\acs{hc}}} \\ \cline{2-4}
                    & n      & 934                  & 934                 \\ \cline{2-4}
                    \multirow{2}{*}{\textbf{Sex}}     & female & 556                  & 548                 \\
                    & male   & 378                  & 368                 \\ \cline{2-4}
                    \multirow{3}{*}{\textbf{Scanner}} & 1      & 285                  & 512                 \\
                    & 2      & 395                  & 236                 \\
                    & 3      & 254                  & 186                 \\
                \end{tabular}
                \subcaption{Summary of sample size (n), sex and scanner site.}
            \end{center}
        \end{subtable}

        \vspace{5mm}

        \begin{subtable}[c]{0.8\textwidth}
            \begin{center}
                \begin{tabular}{rr|SSSS}
                    &                     & {\textbf{Mean}} & {\textbf{SD}} & {\textbf{Min}} & {\textbf{Max}} \\ \cline{2-6}
                    \multirow{2}{*}{\textbf{\acs{mdd}}} & Age ($n=934$)       & 37.76           & 12.94         & 16             & 65             \\
                    & \acs{tiv} ($n=933$) & 1575.94         & 152.97        & 1115.00        & 2189.00        \\  \cline{2-6}
                    \multirow{2}{*}{\textbf{\acs{hc}}}  & Age ($n=934$)       & 34.15           & 12.42         & 17             & 65             \\
                    & \acs{tiv} ($n=933$) & 1575.61         & 156.38        & 1146.23        & 2706.76        \\
                \end{tabular}
                \subcaption{Summage of Age (in years) and \acs{tiv} (\acl{tiv}).}
            \end{center}
        \end{subtable}

        \vspace{6pt}
        \caption[Summary statistics for the final study sample.]{Summary statistics for the final study sample. \\
        \footnotesize
        \setlength{\tabcolsep}{4pt}
        \renewcommand{\arraystretch}{1.0}
            \begin{tabular}{llll}
                \hspace{-6pt} Abbreviations: & \acs{hc}  & - & \acl{hc}  \\
                & \acs{mdd} & - & \acl{mdd} \\
            \end{tabular}
        }
    \end{center}
\end{table}

\FloatBarrier

\section{Predictive Analytics Competition 2018}
\label{cha:pac-2018}
The medical machine learning lab from Prof. Dr. Tim Hahn invited teams from all over the world to develop a model classifying patients suffering from \ac{mdd} and \ac{hc} individuals based on structural Magnetic Resonance Imaging (sMRI) data. This competition was called \emph{Predictive Analytics Competition 2018} (\acs{pac} 2018).

\subsection*{Data}
The data for this competition - comprising structured \ac{mri} datasets with and without MDD from $N\,=\,\num{2240}$ subjects - was provided by the \emph{Institute of Translational Psychiatry}, Münster.
The images were preprocessed in advanced with the SPM toolbox CAT-12  (Matlab 9.0 / SMP12 rev. 6685 / CAT12 v.1184)  and quality checked. The participants got, additional to the diagnosis, the age, gender, \ac{tiv} and the scanner-site to consider them as covariates.

The data was split into a training and test set in advance (Table~\ref{tab:pac_datasplit}). The test set was held back and only used in the final step to designate the winner.

\begin{table}[!ht]
    \renewcommand{\arraystretch}{1.4}

    \begin{center}
        \begin{tabular}{rS[table-format=4.0]S[table-format=3.0]}
            \multicolumn{1}{c}{} & \textbf{Training} & \textbf{Test} \\
            \acs{mdd}            & 1033              & 273           \\
            \acs{hc}             & 759               & 175           \\ \cline{2-3}
            \textbf{Total}       & 1792              & 448
        \end{tabular}
        \vspace{6pt}
        \caption[Split between training and test data for the \acs{pac} 2018.]{Split between training and test data for the \ac{pac} 2018. \\

        \vspace{-6pt}
        \setlength{\tabcolsep}{4pt}
        \renewcommand{\arraystretch}{1.0}
            \begin{tabular}{llll}
                \hspace{-6pt} Abbreviations: & \acs{hc}  & - & \acl{hc}  \\
                & \acs{mdd} & - & \acl{mdd} \\
            \end{tabular}
            \label{tab:pac_datasplit}
        }
    \end{center}
\end{table}

\subsection*{Timeline}

\begin{table}[H]
    \renewcommand{\arraystretch}{1.4}
    \begin{center}
        \begin{tabular}{rrll}
            Registration opened:           & \nth{1}  & February & 2018 \\
            Training data available:       & \nth{1}  & March    & 2018 \\
            Test data available:           & \nth{27} & April    & 2018 \\
            Test prediction upload opened: & \nth{4}  & May      & 2018 \\
            Test prediction upload closed: & \nth{31} & May      & 2018 \\
        \end{tabular}
    \end{center}
\end{table}

The \enquote{\ac{pac} Award Winner 2018} was announced in a ceremony held at annually meeting of the \emph{Organization for Human Brain Mapping} (OHBM) in Singapore at the \nth{21} June 2018.

\subsection*{Results}
In total 49 teams registered with at least 170 participants. The winner was determined by the highest balanced accuracy score on the held-back test set. The team with the highest score was \enquote{paranoidandroid} (Table\ref{tab:top5pac2018}) and won the \ac{pac} 2018 Award.

\begin{table}[!h]
    \renewcommand{\arraystretch}{1.4}
    \begin{center}
        \begin{tabular}{clSSS}
            & \multicolumn{1}{l}{\textbf{Team}} & \textbf{BA} & \textbf{TPR} & \textbf{TNR} \\ \cline{2-5}
            \nth{1} & paranoidandroid                   & 0.65        & 0.53         & 0.77         \\
            \nth{2} & utastoot                          & 0.64        & 0.51         & 0.77         \\
            \nth{3} & berlin\_brain\_decoders           & 0.64        & 0.58         & 0.69         \\
            \nth{4} & depression                        & 0.63        & 0.59         & 0.67         \\
            \nth{5} & neuronauts\_p                     & 0.63        & 0.61         & 0.64
        \end{tabular}
        \vspace{6pt}
        \caption[Test results on the held-back test set by the top five teams at the \ac{pac} 2018.]{Test results on the held-back test set by the top five teams at the \ac{pac} 2018. \\

        \vspace{-6pt}
        \setlength{\tabcolsep}{4pt}
        \renewcommand{\arraystretch}{1.0}
            \begin{tabular}{llll}
                \hspace{-6pt} Abbreviations: & BA  & - & balanced accuracy                \\
                & TPR & - & true positive rate (sensitivity) \\
                & TNR & - & true negative rate (specificity) \\
            \end{tabular}

        }
        \label{tab:top5pac2018}
    \end{center}
\end{table}

\FloatBarrier

\section{Additional tabular presentations}
This section contains the tabular presentation of the graphical visualized data in the paper.

\begin{table}[t]
    \begin{center}
        \begin{subtable}[c]{\textwidth}
            \begin{center}
                \begin{tabular}{rcccccccccc}
                    & & \multicolumn{9}{c}{\textbf{$\geq$ accuracy (\%)}} \\
                    & \multicolumn{1}{c|}{$n$} & 50 & 55 & 60 & 65 & 70 & 75 & 80 & 85 & 90  \\ \cline{2-11}
                    \multirow{12}{*}{\rotatebox[origin=c]{90}{\textbf{overall sample size}}}
                                            & \multicolumn{1}{c|}{10}  & \num{0.54}  & \num{0.54}  & \num{0.43}  & \num{0.33}  & \num{0.15}  & \num{0.15}  & \num{0.05}  & \num{0.05}  & \num{0.01}  \\
                                            & \multicolumn{1}{c|}{20}  & \num{0.64}  & \num{0.52}  & \num{0.40}  & \num{0.24}  & \num{0.13}  & \num{0.06}  & \num{0.01}  & \num{0.01}  & \num{0.00}  \\
                                            & \multicolumn{1}{c|}{30}  & \num{0.70}  & \num{0.61}  & \num{0.39}  & \num{0.29}  & \num{0.11}  & \num{0.06}  & \num{0.01}  & \num{0.00}  & \num{0.00}  \\
                                            & \multicolumn{1}{c|}{40}  & \num{0.78}  & \num{0.63}  & \num{0.47}  & \num{0.23}  & \num{0.09}  & \num{0.02}  & \num{0.00}  & \num{0.00}  & \num{0.00}  \\
                                            & \multicolumn{1}{c|}{50}  & \num{0.82}  & \num{0.70}  & \num{0.51}  & \num{0.27}  & \num{0.08}  & \num{0.02}  & \num{0.00}  & \num{0.00}  & \num{0.00}  \\
                                            & \multicolumn{1}{c|}{60}  & \num{0.81}  & \num{0.62}  & \num{0.41}  & \num{0.19}  & \num{0.05}  & \num{0.01}  & \num{0.00}  & \num{0.00}  & \num{0.00}  \\
                                            & \multicolumn{1}{c|}{70}  & \num{0.88}  & \num{0.72}  & \num{0.42}  & \num{0.20}  & \num{0.07}  & \num{0.01}  & \num{0.00}  & \num{0.00}  & \num{0.00}  \\
                                            & \multicolumn{1}{c|}{80}  & \num{0.90}  & \num{0.73}  & \num{0.48}  & \num{0.18}  & \num{0.04}  & \num{0.00}  & \num{0.00}  & \num{0.00}  & \num{0.00}  \\
                                            & \multicolumn{1}{c|}{90}  & \num{0.89}  & \num{0.76}  & \num{0.45}  & \num{0.20}  & \num{0.03}  & \num{0.00}  & \num{0.00}  & \num{0.00}  & \num{0.00}  \\
                                            & \multicolumn{1}{c|}{100}  & \num{0.93}  & \num{0.75}  & \num{0.47}  & \num{0.17}  & \num{0.02}  & \num{0.00}  & \num{0.00}  & \num{0.00}  & \num{0.00}  \\
                                            & \multicolumn{1}{c|}{125}  & \num{0.94}  & \num{0.81}  & \num{0.48}  & \num{0.14}  & \num{0.01}  & \num{0.00}  & \num{0.00}  & \num{0.00}  & \num{0.00}  \\
                                            & \multicolumn{1}{c|}{150}  & \num{0.96}  & \num{0.83}  & \num{0.45}  & \num{0.12}  & \num{0.01}  & \num{0.00}  & \num{0.00}  & \num{0.00}  & \num{0.00}  \\
                                    \end{tabular}
                \subcaption{Probabilities to yield an accuracy exceeding a certain threshold related to the overall sample size.}
            \end{center}
        \end{subtable}

        \vspace{5mm}

        \begin{subtable}[c]{0.8\textwidth}
            \begin{center}
                \begin{tabular}{rc|ccccc}
                    & \textbf{$n$} & \textbf{mean} & \textbf{median} & \textbf{min} & \textbf{max} & \textbf{std} \\ \cline{2-7}
                    \multirow{12}{*}{\rotatebox[origin=c]{90}{\textbf{overall sample size}}}
                                            & \multicolumn{1}{c|}{10}  & \num{0.55}  & \num{0.60}  & \num{0.00}  & \num{1.00}  & \num{0.20}  \\
                                            & \multicolumn{1}{c|}{20}  & \num{0.56}  & \num{0.60}  & \num{0.10}  & \num{0.95}  & \num{0.15}  \\
                                            & \multicolumn{1}{c|}{30}  & \num{0.58}  & \num{0.58}  & \num{0.20}  & \num{0.87}  & \num{0.12}  \\
                                            & \multicolumn{1}{c|}{40}  & \num{0.59}  & \num{0.60}  & \num{0.25}  & \num{0.82}  & \num{0.10}  \\
                                            & \multicolumn{1}{c|}{50}  & \num{0.59}  & \num{0.60}  & \num{0.16}  & \num{0.82}  & \num{0.09}  \\
                                            & \multicolumn{1}{c|}{60}  & \num{0.58}  & \num{0.58}  & \num{0.32}  & \num{0.82}  & \num{0.08}  \\
                                            & \multicolumn{1}{c|}{70}  & \num{0.59}  & \num{0.60}  & \num{0.36}  & \num{0.81}  & \num{0.07}  \\
                                            & \multicolumn{1}{c|}{80}  & \num{0.59}  & \num{0.60}  & \num{0.36}  & \num{0.76}  & \num{0.07}  \\
                                            & \multicolumn{1}{c|}{90}  & \num{0.59}  & \num{0.60}  & \num{0.34}  & \num{0.77}  & \num{0.07}  \\
                                            & \multicolumn{1}{c|}{100}  & \num{0.60}  & \num{0.60}  & \num{0.35}  & \num{0.81}  & \num{0.06}  \\
                                            & \multicolumn{1}{c|}{125}  & \num{0.59}  & \num{0.60}  & \num{0.41}  & \num{0.74}  & \num{0.05}  \\
                                            & \multicolumn{1}{c|}{150}  & \num{0.59}  & \num{0.59}  & \num{0.41}  & \num{0.73}  & \num{0.05}  \\
                                    \end{tabular}
                \subcaption{Summary of the accuracy value distributions for certain overall sample size.}
            \end{center}
        \end{subtable}

        \caption[Varying overall set size with a linear SVM and LOOCV]{Effects of varying overall sample sizes for training and testing a linear SVM employing LOOCV.}
        \label{tab:overall_linear_svm_loocv}

    \end{center}
\end{table}

\begin{table}[t]
    \begin{center}
        \begin{subtable}[c]{\textwidth}
            \begin{center}
                \begin{tabular}{rcccccccccc}
                    & & \multicolumn{9}{c}{\textbf{$\geq$ accuracy (\%)}} \\
                    & \multicolumn{1}{c|}{$n$} & 50 & 55 & 60 & 65 & 70 & 75 & 80 & 85 & 90  \\ \cline{2-11}
                    \multirow{12}{*}{\rotatebox[origin=c]{90}{\textbf{overall sample size}}}
                                            & \multicolumn{1}{c|}{10}  & \num{0.25}  & \num{0.25}  & \num{0.17}  & \num{0.10}  & \num{0.03}  & \num{0.03}  & \num{0.01}  & \num{0.01}  & \num{0.00}  \\
                                            & \multicolumn{1}{c|}{20}  & \num{0.32}  & \num{0.18}  & \num{0.10}  & \num{0.04}  & \num{0.02}  & \num{0.01}  & \num{0.00}  & \num{0.00}  & \num{0.00}  \\
                                            & \multicolumn{1}{c|}{30}  & \num{0.33}  & \num{0.22}  & \num{0.06}  & \num{0.03}  & \num{0.01}  & \num{0.00}  & \num{0.00}  & \num{0.00}  & \num{0.00}  \\
                                            & \multicolumn{1}{c|}{40}  & \num{0.37}  & \num{0.18}  & \num{0.09}  & \num{0.02}  & \num{0.00}  & \num{0.00}  & \num{0.00}  & \num{0.00}  & \num{0.00}  \\
                                            & \multicolumn{1}{c|}{50}  & \num{0.39}  & \num{0.21}  & \num{0.07}  & \num{0.01}  & \num{0.00}  & \num{0.00}  & \num{0.00}  & \num{0.00}  & \num{0.00}  \\
                                            & \multicolumn{1}{c|}{60}  & \num{0.40}  & \num{0.16}  & \num{0.04}  & \num{0.01}  & \num{0.00}  & \num{0.00}  & \num{0.00}  & \num{0.00}  & \num{0.00}  \\
                                            & \multicolumn{1}{c|}{70}  & \num{0.40}  & \num{0.16}  & \num{0.02}  & \num{0.00}  & \num{0.00}  & \num{0.00}  & \num{0.00}  & \num{0.00}  & \num{0.00}  \\
                                            & \multicolumn{1}{c|}{80}  & \num{0.41}  & \num{0.13}  & \num{0.03}  & \num{0.00}  & \num{0.00}  & \num{0.00}  & \num{0.00}  & \num{0.00}  & \num{0.00}  \\
                                            & \multicolumn{1}{c|}{90}  & \num{0.41}  & \num{0.14}  & \num{0.02}  & \num{0.00}  & \num{0.00}  & \num{0.00}  & \num{0.00}  & \num{0.00}  & \num{0.00}  \\
                                            & \multicolumn{1}{c|}{100}  & \num{0.43}  & \num{0.11}  & \num{0.02}  & \num{0.00}  & \num{0.00}  & \num{0.00}  & \num{0.00}  & \num{0.00}  & \num{0.00}  \\
                                            & \multicolumn{1}{c|}{125}  & \num{0.46}  & \num{0.13}  & \num{0.01}  & \num{0.00}  & \num{0.00}  & \num{0.00}  & \num{0.00}  & \num{0.00}  & \num{0.00}  \\
                                            & \multicolumn{1}{c|}{150}  & \num{0.44}  & \num{0.09}  & \num{0.01}  & \num{0.00}  & \num{0.00}  & \num{0.00}  & \num{0.00}  & \num{0.00}  & \num{0.00}  \\
                                    \end{tabular}
                \subcaption{Probabilities to yield an accuracy exceeding a certain threshold related to the overall sample size.}
            \end{center}
        \end{subtable}

        \vspace{5mm}

        \begin{subtable}[c]{0.8\textwidth}
            \begin{center}
                \begin{tabular}{rc|ccccc}
                    & \textbf{$n$} & \textbf{mean} & \textbf{median} & \textbf{min} & \textbf{max} & \textbf{std} \\ \cline{2-7}
                    \multirow{12}{*}{\rotatebox[origin=c]{90}{\textbf{overall sample size}}}
                                            & \multicolumn{1}{c|}{10}  & \num{0.44}  & \num{0.40}  & \num{0.00}  & \num{1.00}  & \num{0.17}  \\
                                            & \multicolumn{1}{c|}{20}  & \num{0.47}  & \num{0.45}  & \num{0.15}  & \num{0.80}  & \num{0.11}  \\
                                            & \multicolumn{1}{c|}{30}  & \num{0.48}  & \num{0.47}  & \num{0.23}  & \num{0.80}  & \num{0.09}  \\
                                            & \multicolumn{1}{c|}{40}  & \num{0.49}  & \num{0.50}  & \num{0.25}  & \num{0.70}  & \num{0.08}  \\
                                            & \multicolumn{1}{c|}{50}  & \num{0.49}  & \num{0.48}  & \num{0.28}  & \num{0.74}  & \num{0.07}  \\
                                            & \multicolumn{1}{c|}{60}  & \num{0.49}  & \num{0.50}  & \num{0.28}  & \num{0.68}  & \num{0.06}  \\
                                            & \multicolumn{1}{c|}{70}  & \num{0.49}  & \num{0.50}  & \num{0.31}  & \num{0.69}  & \num{0.06}  \\
                                            & \multicolumn{1}{c|}{80}  & \num{0.49}  & \num{0.50}  & \num{0.30}  & \num{0.66}  & \num{0.06}  \\
                                            & \multicolumn{1}{c|}{90}  & \num{0.49}  & \num{0.49}  & \num{0.34}  & \num{0.68}  & \num{0.05}  \\
                                            & \multicolumn{1}{c|}{100}  & \num{0.50}  & \num{0.50}  & \num{0.33}  & \num{0.64}  & \num{0.05}  \\
                                            & \multicolumn{1}{c|}{125}  & \num{0.50}  & \num{0.50}  & \num{0.33}  & \num{0.63}  & \num{0.05}  \\
                                            & \multicolumn{1}{c|}{150}  & \num{0.50}  & \num{0.49}  & \num{0.37}  & \num{0.63}  & \num{0.04}  \\
                                    \end{tabular}
                \subcaption{Summary of the accuracy value distributions for certain overall sample size.}
            \end{center}
        \end{subtable}

        \caption[Varying overall set size with a Dummy Classifier and LOOCV]{Effects of varying overall sample sizes for training and testing a Dummy Classifier employing LOOCV.}
        \label{tab:overall_dummy_classifier_loocv}

    \end{center}
\end{table}

\begin{table}[t]
    \begin{center}
        \begin{subtable}[c]{\textwidth}
            \begin{center}
                \begin{tabular}{rcccccccccc}
                    & & \multicolumn{9}{c}{\textbf{$\geq$ accuracy (\%)}} \\
                    & \multicolumn{1}{c|}{$n$} & 50 & 55 & 60 & 65 & 70 & 75 & 80 & 85 & 90  \\ \cline{2-11}
                    \multirow{12}{*}{\rotatebox[origin=c]{90}{\textbf{train sample size}}}
                                            & \multicolumn{1}{c|}{10}  & \num{0.80}  & \num{0.62}  & \num{0.42}  & \num{0.16}  & \num{0.00}  & \num{0.00}  & \num{0.00}  & \num{0.00}  & \num{0.00}  \\
                                            & \multicolumn{1}{c|}{20}  & \num{0.87}  & \num{0.75}  & \num{0.55}  & \num{0.20}  & \num{0.00}  & \num{0.00}  & \num{0.00}  & \num{0.00}  & \num{0.00}  \\
                                            & \multicolumn{1}{c|}{30}  & \num{0.94}  & \num{0.83}  & \num{0.65}  & \num{0.26}  & \num{0.00}  & \num{0.00}  & \num{0.00}  & \num{0.00}  & \num{0.00}  \\
                                            & \multicolumn{1}{c|}{40}  & \num{0.97}  & \num{0.90}  & \num{0.72}  & \num{0.29}  & \num{0.00}  & \num{0.00}  & \num{0.00}  & \num{0.00}  & \num{0.00}  \\
                                            & \multicolumn{1}{c|}{50}  & \num{0.97}  & \num{0.93}  & \num{0.76}  & \num{0.31}  & \num{0.00}  & \num{0.00}  & \num{0.00}  & \num{0.00}  & \num{0.00}  \\
                                            & \multicolumn{1}{c|}{60}  & \num{0.99}  & \num{0.95}  & \num{0.76}  & \num{0.30}  & \num{0.00}  & \num{0.00}  & \num{0.00}  & \num{0.00}  & \num{0.00}  \\
                                            & \multicolumn{1}{c|}{70}  & \num{0.99}  & \num{0.96}  & \num{0.83}  & \num{0.35}  & \num{0.00}  & \num{0.00}  & \num{0.00}  & \num{0.00}  & \num{0.00}  \\
                                            & \multicolumn{1}{c|}{80}  & \num{0.99}  & \num{0.97}  & \num{0.83}  & \num{0.34}  & \num{0.00}  & \num{0.00}  & \num{0.00}  & \num{0.00}  & \num{0.00}  \\
                                            & \multicolumn{1}{c|}{90}  & \num{1.00}  & \num{0.98}  & \num{0.89}  & \num{0.36}  & \num{0.00}  & \num{0.00}  & \num{0.00}  & \num{0.00}  & \num{0.00}  \\
                                            & \multicolumn{1}{c|}{100}  & \num{1.00}  & \num{0.98}  & \num{0.87}  & \num{0.35}  & \num{0.00}  & \num{0.00}  & \num{0.00}  & \num{0.00}  & \num{0.00}  \\
                                            & \multicolumn{1}{c|}{125}  & \num{1.00}  & \num{0.99}  & \num{0.91}  & \num{0.39}  & \num{0.00}  & \num{0.00}  & \num{0.00}  & \num{0.00}  & \num{0.00}  \\
                                            & \multicolumn{1}{c|}{150}  & \num{1.00}  & \num{0.99}  & \num{0.94}  & \num{0.39}  & \num{0.00}  & \num{0.00}  & \num{0.00}  & \num{0.00}  & \num{0.00}  \\
                                    \end{tabular}
                \subcaption{Probabilities to yield an accuracy exceeding a certain threshold related to the train sample size.}
            \end{center}
        \end{subtable}

        \vspace{5mm}

        \begin{subtable}[c]{0.8\textwidth}
            \begin{center}
                \begin{tabular}{rc|ccccc}
                    & \textbf{$n$} & \textbf{mean} & \textbf{median} & \textbf{min} & \textbf{max} & \textbf{std} \\ \cline{2-7}
                    \multirow{12}{*}{\rotatebox[origin=c]{90}{\textbf{train sample size}}}
                                            & \multicolumn{1}{c|}{10}  & \num{0.57}  & \num{0.58}  & \num{0.33}  & \num{0.69}  & \num{0.08}  \\
                                            & \multicolumn{1}{c|}{20}  & \num{0.59}  & \num{0.61}  & \num{0.32}  & \num{0.69}  & \num{0.07}  \\
                                            & \multicolumn{1}{c|}{30}  & \num{0.61}  & \num{0.62}  & \num{0.36}  & \num{0.69}  & \num{0.06}  \\
                                            & \multicolumn{1}{c|}{40}  & \num{0.62}  & \num{0.63}  & \num{0.35}  & \num{0.70}  & \num{0.05}  \\
                                            & \multicolumn{1}{c|}{50}  & \num{0.62}  & \num{0.63}  & \num{0.38}  & \num{0.70}  & \num{0.05}  \\
                                            & \multicolumn{1}{c|}{60}  & \num{0.63}  & \num{0.64}  & \num{0.46}  & \num{0.70}  & \num{0.04}  \\
                                            & \multicolumn{1}{c|}{70}  & \num{0.63}  & \num{0.64}  & \num{0.45}  & \num{0.71}  & \num{0.04}  \\
                                            & \multicolumn{1}{c|}{80}  & \num{0.63}  & \num{0.64}  & \num{0.45}  & \num{0.70}  & \num{0.03}  \\
                                            & \multicolumn{1}{c|}{90}  & \num{0.64}  & \num{0.64}  & \num{0.49}  & \num{0.71}  & \num{0.03}  \\
                                            & \multicolumn{1}{c|}{100}  & \num{0.64}  & \num{0.64}  & \num{0.51}  & \num{0.70}  & \num{0.03}  \\
                                            & \multicolumn{1}{c|}{125}  & \num{0.64}  & \num{0.64}  & \num{0.51}  & \num{0.71}  & \num{0.03}  \\
                                            & \multicolumn{1}{c|}{150}  & \num{0.64}  & \num{0.65}  & \num{0.51}  & \num{0.70}  & \num{0.03}  \\
                                    \end{tabular}
                \subcaption{Summary of the accuracy value distributions for certain train sample size.}
            \end{center}
        \end{subtable}

        \caption[Varying train set size and fixed validation set on a linear SVM]{Results of a fixed validation set of $n=300$ and varying train sample sizes on a linear SVM using default parameter. ($C=1$)}
        \label{tab:train_set_linear_svm}

    \end{center}
\end{table}

\begin{table}[t]
    \begin{center}
        \begin{subtable}[c]{\textwidth}
            \begin{center}
                \begin{tabular}{rcccccccccc}
                    & & \multicolumn{9}{c}{\textbf{$\geq$ accuracy (\%)}} \\
                    & \multicolumn{1}{c|}{$n$} & 50 & 55 & 60 & 65 & 70 & 75 & 80 & 85 & 90  \\ \cline{2-11}
                    \multirow{12}{*}{\rotatebox[origin=c]{90}{\textbf{train sample size}}}
                                            & \multicolumn{1}{c|}{10}  & \num{0.49}  & \num{0.04}  & \num{0.00}  & \num{0.00}  & \num{0.00}  & \num{0.00}  & \num{0.00}  & \num{0.00}  & \num{0.00}  \\
                                            & \multicolumn{1}{c|}{20}  & \num{0.47}  & \num{0.04}  & \num{0.00}  & \num{0.00}  & \num{0.00}  & \num{0.00}  & \num{0.00}  & \num{0.00}  & \num{0.00}  \\
                                            & \multicolumn{1}{c|}{30}  & \num{0.47}  & \num{0.04}  & \num{0.00}  & \num{0.00}  & \num{0.00}  & \num{0.00}  & \num{0.00}  & \num{0.00}  & \num{0.00}  \\
                                            & \multicolumn{1}{c|}{40}  & \num{0.47}  & \num{0.03}  & \num{0.00}  & \num{0.00}  & \num{0.00}  & \num{0.00}  & \num{0.00}  & \num{0.00}  & \num{0.00}  \\
                                            & \multicolumn{1}{c|}{50}  & \num{0.45}  & \num{0.04}  & \num{0.00}  & \num{0.00}  & \num{0.00}  & \num{0.00}  & \num{0.00}  & \num{0.00}  & \num{0.00}  \\
                                            & \multicolumn{1}{c|}{60}  & \num{0.46}  & \num{0.04}  & \num{0.00}  & \num{0.00}  & \num{0.00}  & \num{0.00}  & \num{0.00}  & \num{0.00}  & \num{0.00}  \\
                                            & \multicolumn{1}{c|}{70}  & \num{0.48}  & \num{0.03}  & \num{0.00}  & \num{0.00}  & \num{0.00}  & \num{0.00}  & \num{0.00}  & \num{0.00}  & \num{0.00}  \\
                                            & \multicolumn{1}{c|}{80}  & \num{0.49}  & \num{0.04}  & \num{0.00}  & \num{0.00}  & \num{0.00}  & \num{0.00}  & \num{0.00}  & \num{0.00}  & \num{0.00}  \\
                                            & \multicolumn{1}{c|}{90}  & \num{0.47}  & \num{0.03}  & \num{0.00}  & \num{0.00}  & \num{0.00}  & \num{0.00}  & \num{0.00}  & \num{0.00}  & \num{0.00}  \\
                                            & \multicolumn{1}{c|}{100}  & \num{0.48}  & \num{0.05}  & \num{0.00}  & \num{0.00}  & \num{0.00}  & \num{0.00}  & \num{0.00}  & \num{0.00}  & \num{0.00}  \\
                                            & \multicolumn{1}{c|}{125}  & \num{0.45}  & \num{0.03}  & \num{0.00}  & \num{0.00}  & \num{0.00}  & \num{0.00}  & \num{0.00}  & \num{0.00}  & \num{0.00}  \\
                                            & \multicolumn{1}{c|}{150}  & \num{0.45}  & \num{0.03}  & \num{0.00}  & \num{0.00}  & \num{0.00}  & \num{0.00}  & \num{0.00}  & \num{0.00}  & \num{0.00}  \\
                                    \end{tabular}
                \subcaption{Probabilities to yield an accuracy exceeding a certain threshold related to the train sample size.}
            \end{center}
        \end{subtable}

        \vspace{5mm}

        \begin{subtable}[c]{0.8\textwidth}
            \begin{center}
                \begin{tabular}{rc|ccccc}
                    & \textbf{$n$} & \textbf{mean} & \textbf{median} & \textbf{min} & \textbf{max} & \textbf{std} \\ \cline{2-7}
                    \multirow{12}{*}{\rotatebox[origin=c]{90}{\textbf{train sample size}}}
                                            & \multicolumn{1}{c|}{10}  & \num{0.50}  & \num{0.50}  & \num{0.41}  & \num{0.59}  & \num{0.03}  \\
                                            & \multicolumn{1}{c|}{20}  & \num{0.50}  & \num{0.50}  & \num{0.39}  & \num{0.59}  & \num{0.03}  \\
                                            & \multicolumn{1}{c|}{30}  & \num{0.50}  & \num{0.50}  & \num{0.41}  & \num{0.58}  & \num{0.03}  \\
                                            & \multicolumn{1}{c|}{40}  & \num{0.50}  & \num{0.50}  & \num{0.38}  & \num{0.60}  & \num{0.03}  \\
                                            & \multicolumn{1}{c|}{50}  & \num{0.50}  & \num{0.50}  & \num{0.42}  & \num{0.61}  & \num{0.03}  \\
                                            & \multicolumn{1}{c|}{60}  & \num{0.50}  & \num{0.50}  & \num{0.40}  & \num{0.60}  & \num{0.03}  \\
                                            & \multicolumn{1}{c|}{70}  & \num{0.50}  & \num{0.50}  & \num{0.41}  & \num{0.58}  & \num{0.03}  \\
                                            & \multicolumn{1}{c|}{80}  & \num{0.50}  & \num{0.50}  & \num{0.42}  & \num{0.62}  & \num{0.03}  \\
                                            & \multicolumn{1}{c|}{90}  & \num{0.50}  & \num{0.50}  & \num{0.42}  & \num{0.58}  & \num{0.03}  \\
                                            & \multicolumn{1}{c|}{100}  & \num{0.50}  & \num{0.50}  & \num{0.40}  & \num{0.60}  & \num{0.03}  \\
                                            & \multicolumn{1}{c|}{125}  & \num{0.50}  & \num{0.50}  & \num{0.40}  & \num{0.59}  & \num{0.03}  \\
                                            & \multicolumn{1}{c|}{150}  & \num{0.50}  & \num{0.50}  & \num{0.39}  & \num{0.61}  & \num{0.03}  \\
                                    \end{tabular}
                \subcaption{Summary of the accuracy value distributions for certain train sample size.}
            \end{center}
        \end{subtable}

        \caption[Varying train set size and fixed validation set on a Dummy Classifier]{Results of a fixed validation set of $n=300$ and varying train sample sizes on a Dummy Classifier.}
        \label{tab:train_dummy_classifier}

    \end{center}
\end{table}

\begin{table}[t]
    \begin{center}
        \begin{subtable}[c]{\textwidth}
            \begin{center}
                \begin{tabular}{rcccccccccc}
                    & & \multicolumn{9}{c}{\textbf{$\geq$ accuracy (\%)}} \\
                    & \multicolumn{1}{c|}{$n$} & 50 & 55 & 60 & 65 & 70 & 75 & 80 & 85 & 90  \\ \cline{2-11}
                    \multirow{12}{*}{\rotatebox[origin=c]{90}{\textbf{test sample size}}}
                                            & \multicolumn{1}{c|}{10}  & \num{0.78}  & \num{0.78}  & \num{0.66}  & \num{0.54}  & \num{0.29}  & \num{0.29}  & \num{0.10}  & \num{0.10}  & \num{0.02}  \\
                                            & \multicolumn{1}{c|}{20}  & \num{0.91}  & \num{0.81}  & \num{0.69}  & \num{0.48}  & \num{0.30}  & \num{0.17}  & \num{0.06}  & \num{0.06}  & \num{0.00}  \\
                                            & \multicolumn{1}{c|}{30}  & \num{0.96}  & \num{0.90}  & \num{0.73}  & \num{0.57}  & \num{0.27}  & \num{0.15}  & \num{0.02}  & \num{0.01}  & \num{0.00}  \\
                                            & \multicolumn{1}{c|}{40}  & \num{0.98}  & \num{0.91}  & \num{0.80}  & \num{0.49}  & \num{0.25}  & \num{0.09}  & \num{0.02}  & \num{0.01}  & \num{0.00}  \\
                                            & \multicolumn{1}{c|}{50}  & \num{0.99}  & \num{0.97}  & \num{0.85}  & \num{0.58}  & \num{0.22}  & \num{0.07}  & \num{0.00}  & \num{0.00}  & \num{0.00}  \\
                                            & \multicolumn{1}{c|}{60}  & \num{1.00}  & \num{0.97}  & \num{0.84}  & \num{0.56}  & \num{0.23}  & \num{0.05}  & \num{0.00}  & \num{0.00}  & \num{0.00}  \\
                                            & \multicolumn{1}{c|}{70}  & \num{1.00}  & \num{0.99}  & \num{0.88}  & \num{0.61}  & \num{0.21}  & \num{0.04}  & \num{0.00}  & \num{0.00}  & \num{0.00}  \\
                                            & \multicolumn{1}{c|}{80}  & \num{1.00}  & \num{0.99}  & \num{0.90}  & \num{0.56}  & \num{0.17}  & \num{0.02}  & \num{0.00}  & \num{0.00}  & \num{0.00}  \\
                                            & \multicolumn{1}{c|}{90}  & \num{1.00}  & \num{1.00}  & \num{0.94}  & \num{0.63}  & \num{0.17}  & \num{0.02}  & \num{0.00}  & \num{0.00}  & \num{0.00}  \\
                                            & \multicolumn{1}{c|}{100}  & \num{1.00}  & \num{0.99}  & \num{0.94}  & \num{0.58}  & \num{0.13}  & \num{0.01}  & \num{0.00}  & \num{0.00}  & \num{0.00}  \\
                                            & \multicolumn{1}{c|}{125}  & \num{1.00}  & \num{1.00}  & \num{0.97}  & \num{0.63}  & \num{0.13}  & \num{0.00}  & \num{0.00}  & \num{0.00}  & \num{0.00}  \\
                                            & \multicolumn{1}{c|}{150}  & \num{1.00}  & \num{1.00}  & \num{0.98}  & \num{0.66}  & \num{0.08}  & \num{0.00}  & \num{0.00}  & \num{0.00}  & \num{0.00}  \\
                                    \end{tabular}
                \subcaption{Probabilities to yield an accuracy exceeding a certain threshold related to the test sample size.}
            \end{center}
        \end{subtable}

        \vspace{5mm}

        \begin{subtable}[c]{0.8\textwidth}
            \begin{center}
                \begin{tabular}{rc|ccccc}
                    & \textbf{$n$} & \textbf{mean} & \textbf{median} & \textbf{min} & \textbf{max} & \textbf{std} \\ \cline{2-7}
                    \multirow{12}{*}{\rotatebox[origin=c]{90}{\textbf{test sample size}}}
                                            & \multicolumn{1}{c|}{10}  & \num{0.66}  & \num{0.70}  & \num{0.20}  & \num{1.00}  & \num{0.15}  \\
                                            & \multicolumn{1}{c|}{20}  & \num{0.67}  & \num{0.65}  & \num{0.35}  & \num{0.95}  & \num{0.11}  \\
                                            & \multicolumn{1}{c|}{30}  & \num{0.66}  & \num{0.67}  & \num{0.40}  & \num{0.93}  & \num{0.08}  \\
                                            & \multicolumn{1}{c|}{40}  & \num{0.66}  & \num{0.65}  & \num{0.42}  & \num{0.88}  & \num{0.07}  \\
                                            & \multicolumn{1}{c|}{50}  & \num{0.66}  & \num{0.66}  & \num{0.48}  & \num{0.88}  & \num{0.06}  \\
                                            & \multicolumn{1}{c|}{60}  & \num{0.66}  & \num{0.67}  & \num{0.47}  & \num{0.83}  & \num{0.06}  \\
                                            & \multicolumn{1}{c|}{70}  & \num{0.66}  & \num{0.67}  & \num{0.50}  & \num{0.83}  & \num{0.05}  \\
                                            & \multicolumn{1}{c|}{80}  & \num{0.66}  & \num{0.66}  & \num{0.49}  & \num{0.80}  & \num{0.05}  \\
                                            & \multicolumn{1}{c|}{90}  & \num{0.66}  & \num{0.67}  & \num{0.53}  & \num{0.78}  & \num{0.04}  \\
                                            & \multicolumn{1}{c|}{100}  & \num{0.66}  & \num{0.66}  & \num{0.51}  & \num{0.79}  & \num{0.04}  \\
                                            & \multicolumn{1}{c|}{125}  & \num{0.66}  & \num{0.66}  & \num{0.54}  & \num{0.78}  & \num{0.03}  \\
                                            & \multicolumn{1}{c|}{150}  & \num{0.66}  & \num{0.66}  & \num{0.56}  & \num{0.75}  & \num{0.03}  \\
                                    \end{tabular}
                \subcaption{Summary of the accuracy value distributions for certain test sample size.}
            \end{center}
        \end{subtable}

        \caption[Varying test set size on a fixed pretrained linear SVM]{Effects of varying test sample sizes on a fixed pretrained linear SVM.}
        \label{tab:test_set_linear_svm}

    \end{center}
\end{table}

\begin{table}[t]
    \begin{center}
        \begin{subtable}[c]{\textwidth}
            \begin{center}
                \begin{tabular}{rcccccccccc}
                    & & \multicolumn{9}{c}{\textbf{$\geq$ accuracy (\%)}} \\
                    & \multicolumn{1}{c|}{$n$} & 50 & 55 & 60 & 65 & 70 & 75 & 80 & 85 & 90  \\ \cline{2-11}
                    \multirow{12}{*}{\rotatebox[origin=c]{90}{\textbf{test sample size}}}
                                            & \multicolumn{1}{c|}{10}  & \num{0.39}  & \num{0.39}  & \num{0.29}  & \num{0.17}  & \num{0.05}  & \num{0.05}  & \num{0.01}  & \num{0.01}  & \num{0.00}  \\
                                            & \multicolumn{1}{c|}{20}  & \num{0.42}  & \num{0.24}  & \num{0.13}  & \num{0.05}  & \num{0.02}  & \num{0.01}  & \num{0.00}  & \num{0.00}  & \num{0.00}  \\
                                            & \multicolumn{1}{c|}{30}  & \num{0.42}  & \num{0.28}  & \num{0.10}  & \num{0.05}  & \num{0.01}  & \num{0.00}  & \num{0.00}  & \num{0.00}  & \num{0.00}  \\
                                            & \multicolumn{1}{c|}{40}  & \num{0.44}  & \num{0.19}  & \num{0.08}  & \num{0.01}  & \num{0.00}  & \num{0.00}  & \num{0.00}  & \num{0.00}  & \num{0.00}  \\
                                            & \multicolumn{1}{c|}{50}  & \num{0.44}  & \num{0.20}  & \num{0.06}  & \num{0.01}  & \num{0.00}  & \num{0.00}  & \num{0.00}  & \num{0.00}  & \num{0.00}  \\
                                            & \multicolumn{1}{c|}{60}  & \num{0.45}  & \num{0.15}  & \num{0.03}  & \num{0.00}  & \num{0.00}  & \num{0.00}  & \num{0.00}  & \num{0.00}  & \num{0.00}  \\
                                            & \multicolumn{1}{c|}{70}  & \num{0.45}  & \num{0.17}  & \num{0.02}  & \num{0.00}  & \num{0.00}  & \num{0.00}  & \num{0.00}  & \num{0.00}  & \num{0.00}  \\
                                            & \multicolumn{1}{c|}{80}  & \num{0.44}  & \num{0.10}  & \num{0.01}  & \num{0.00}  & \num{0.00}  & \num{0.00}  & \num{0.00}  & \num{0.00}  & \num{0.00}  \\
                                            & \multicolumn{1}{c|}{90}  & \num{0.43}  & \num{0.12}  & \num{0.01}  & \num{0.00}  & \num{0.00}  & \num{0.00}  & \num{0.00}  & \num{0.00}  & \num{0.00}  \\
                                            & \multicolumn{1}{c|}{100}  & \num{0.47}  & \num{0.10}  & \num{0.01}  & \num{0.00}  & \num{0.00}  & \num{0.00}  & \num{0.00}  & \num{0.00}  & \num{0.00}  \\
                                            & \multicolumn{1}{c|}{125}  & \num{0.53}  & \num{0.09}  & \num{0.00}  & \num{0.00}  & \num{0.00}  & \num{0.00}  & \num{0.00}  & \num{0.00}  & \num{0.00}  \\
                                            & \multicolumn{1}{c|}{150}  & \num{0.45}  & \num{0.04}  & \num{0.00}  & \num{0.00}  & \num{0.00}  & \num{0.00}  & \num{0.00}  & \num{0.00}  & \num{0.00}  \\
                                    \end{tabular}
                \subcaption{Probabilities to yield an accuracy exceeding a certain threshold related to the test sample size.}
            \end{center}
        \end{subtable}

        \vspace{5mm}

        \begin{subtable}[c]{0.8\textwidth}
            \begin{center}
                \begin{tabular}{rc|ccccc}
                    & \textbf{$n$} & \textbf{mean} & \textbf{median} & \textbf{min} & \textbf{max} & \textbf{std} \\ \cline{2-7}
                    \multirow{12}{*}{\rotatebox[origin=c]{90}{\textbf{test sample size}}}
                                            & \multicolumn{1}{c|}{10}  & \num{0.50}  & \num{0.50}  & \num{0.10}  & \num{1.00}  & \num{0.16}  \\
                                            & \multicolumn{1}{c|}{20}  & \num{0.50}  & \num{0.50}  & \num{0.15}  & \num{0.85}  & \num{0.11}  \\
                                            & \multicolumn{1}{c|}{30}  & \num{0.50}  & \num{0.50}  & \num{0.23}  & \num{0.73}  & \num{0.09}  \\
                                            & \multicolumn{1}{c|}{40}  & \num{0.50}  & \num{0.50}  & \num{0.28}  & \num{0.72}  & \num{0.07}  \\
                                            & \multicolumn{1}{c|}{50}  & \num{0.50}  & \num{0.50}  & \num{0.32}  & \num{0.70}  & \num{0.06}  \\
                                            & \multicolumn{1}{c|}{60}  & \num{0.50}  & \num{0.50}  & \num{0.30}  & \num{0.67}  & \num{0.06}  \\
                                            & \multicolumn{1}{c|}{70}  & \num{0.50}  & \num{0.50}  & \num{0.34}  & \num{0.67}  & \num{0.05}  \\
                                            & \multicolumn{1}{c|}{80}  & \num{0.50}  & \num{0.50}  & \num{0.35}  & \num{0.64}  & \num{0.05}  \\
                                            & \multicolumn{1}{c|}{90}  & \num{0.50}  & \num{0.50}  & \num{0.36}  & \num{0.67}  & \num{0.04}  \\
                                            & \multicolumn{1}{c|}{100}  & \num{0.50}  & \num{0.50}  & \num{0.39}  & \num{0.63}  & \num{0.04}  \\
                                            & \multicolumn{1}{c|}{125}  & \num{0.50}  & \num{0.50}  & \num{0.39}  & \num{0.62}  & \num{0.04}  \\
                                            & \multicolumn{1}{c|}{150}  & \num{0.50}  & \num{0.50}  & \num{0.40}  & \num{0.59}  & \num{0.03}  \\
                                    \end{tabular}
                \subcaption{Summary of the accuracy value distributions for certain test sample size.}
            \end{center}
        \end{subtable}

        \caption[Varying test set size on a fixed pretrained Dummy Classifier]{Effects of varying test sample sizes on a fixed pretrained Dummy Classifier.}
        \label{tab:test_set_dummy_classifier}

    \end{center}
\end{table}

\FloatBarrier

\section{Influence of the scanner distribution}
\label{cha:influence-of-scanner-distribution}
Scanner site distributions were not well balanced in the \ac{pac} sample.
This imbalance was even stronger in our randomly drawn sub-samples. To determine the influence of the different scanner-sites on model accuracy, we took the same methodical approach that we used for the “train sample size effect analysis” and determined the scanner distribution for each set. Following, we calculated Spearman’s rho between the scanner-distribution and the accuracy of the hold-out test set ($n=300$). The scanner-distribution for each set was approximated using Gini’s index. Here, we show that the scanner distribution has a statistically significant influence on test set accuracy but explains only \SI{0.79}{\percent} of the variance.

\FloatBarrier

\section{Adjustment of SVM regularization based on sample size}
\label{cha:adjustment-svm-regularization}
\begin{figure}[ht]
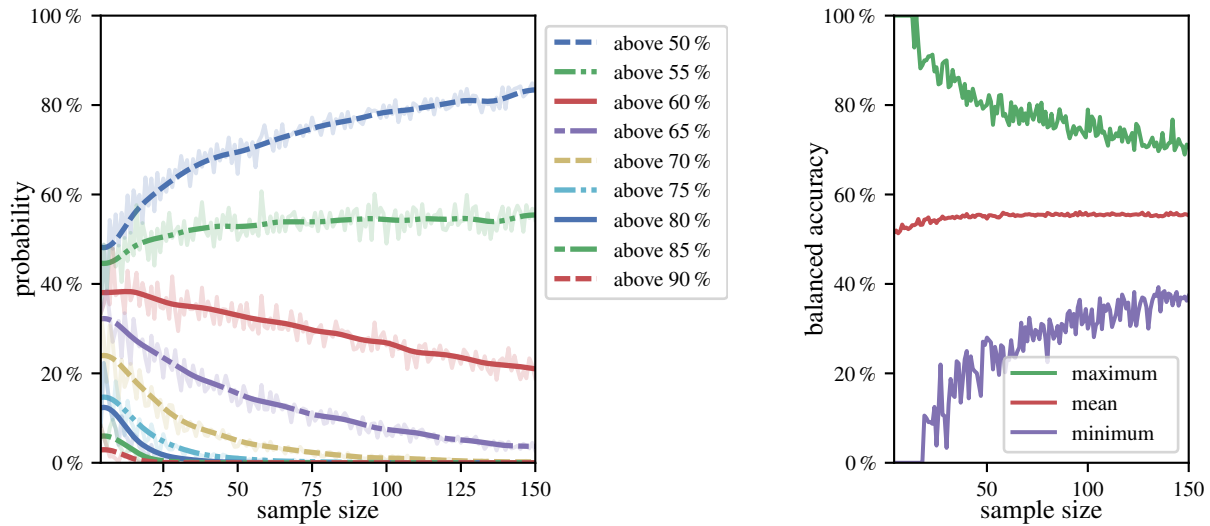

    \captionsetup[subfigure]{justification=justified,singlelinecheck=false}
    \begin{subfigure}[t]{0.61\textwidth}
        \input{images/overall_set_size_svm_adj_c_chances.pgf}
        \caption{Probabilities for linear \acp{svm} to yield an accuracy exceeding a certain threshold as a function of sample size employing \ac{loocv}.}
    \end{subfigure}
    \hspace{3.0mm}
    \begin{subfigure}[t]{0.34\textwidth}
        \input{images/overall_set_size_svm_adj_c_stats.pgf}
        \caption{Minimum, maximum and mean results for the linear \acp{svm} as a function of sample size employing \ac{loocv}. }
    \end{subfigure}
    \caption[Effects of varying overall sample sizes with an adjusted $C$ parameter.]{Effects of varying overall sample sizes employing \ac{loocv} with an adjusted $C$ parameter with $C=\frac{75}{\text{sample size}}$.}
    \label{fig:adj_c_overall}
\end{figure}

\begin{figure}[ht]
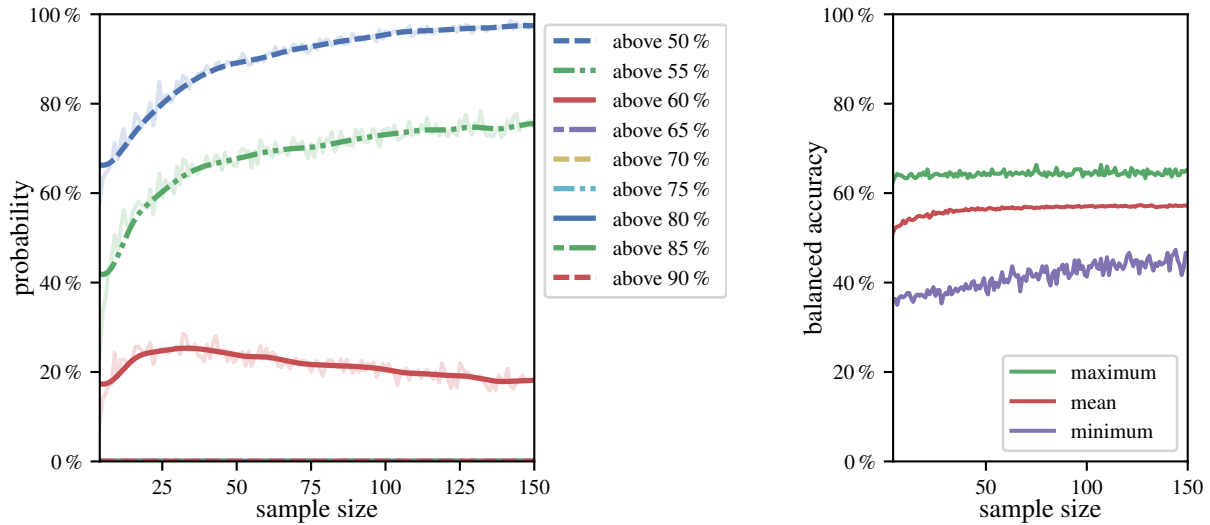

    \captionsetup[subfigure]{justification=justified,singlelinecheck=false}
    \begin{subfigure}[t]{0.61\textwidth}
        \input{images/train_set_size_svm_adj_c_chances.pgf}
        \caption{Probabilities for linear \acp{svm} to yield an accuracy exceeding a certain threshold as a function of training sample size.}
    \end{subfigure}
    \hspace{3.0mm}
    \begin{subfigure}[t]{0.34\textwidth}
        \input{images/train_set_size_svm_adj_c_stats.pgf}
        \caption{Minimum, maximum and mean results for the linear \acp{svm} as a function of training sample size.}
    \end{subfigure}
    \caption[Effects of varying train sample sizes with an adjusted $C$ parameter.]{Results as a function of training set sizes with a fixed test set size of $N = 300$ and an adjusted $C$ parameter with $C=\frac{75}{\text{sample size}}$.}
    \label{fig:adj_c_test}
\end{figure}

The regularization of a linear \ac{svm} depends on the absolute number of outliers in a sample. To exclude the possibility that the use of default hyperparameters (a constant value of $C = 1$) caused the effect that we observed, we have adjusted the \ac{svm} regularization (our $C$ hyperparameter) based on the size of each analyses sub-sample. In this adjustment, a constant value $k$ is divided by the sample size. To approximate the default parameter $C=1$ on average, we have set $k=75$. For example, $C$ would be $C = 100/75 = \num{1.33}$ for a sample size of $n=100$. In this case, the adjusted regularization did not deliver results as good as what we observed with $C=1$ (see Figure~\ref{fig:adj_c_overall},~\ref{fig:adj_c_test}). Therefore, we conclude that the default value of $C=1$ across changing N values did not increase the probability of misestimating accuracy as sample size decreased.

% Supplement Figure 1

% Supplement Figure 2
\FloatBarrier

\section{Alternative Machine Configurations}
\label{cha:alternative-machine-configurations}
To exclude the possibility that the observed effects can only be traced back to our specific configuration, we have tested other usual configurations. The configurations consist of a preprocessing and a classification. In the preprocessing step, we used a method for reduction of feature space and a method for features selection. The reduction of the feature space was achieved with a \ac{pca}, where only a certain number the first components were used. Afterwards, for feature selection, an ANOVA was calculated and a specific number of feature beginning with the highest F-value were taken. The actual configuration used for preprocessing is listed in Table~\ref{tab:preprocessing}.
For the classification, we chose three specific machines:
\begin{enumerate}
    \item An \ac{svm} with a linear kernel and default parameter. ($C = \num{1.0}$)
    \item An \ac{svm} with an \ac{rbf} kernel and default parameter ($C = \num{1.0}$; $\gamma = 1/n_\text{feature}$)
    \item A Random Forrest and default parameter ($n_\text{estimators}=\num{100}$)
\end{enumerate}

In combination with the preprocessing, this results in 48 configurations. Since the found effect is limited to the test set size, we have only repeated our analyses for the test set for each of these configurations.

\begin{table}[!htp]
    \begin{center}

        \begin{tabular}{rr|ccccc}
            & & \multicolumn{5}{c}{\textbf{\acs{pca} $n_\text{components}$}} \\
            &             & no       & all      & \num{500} & \num{50} & \num{10} \\
            \cline{2-7}
            \multirow{6}{*}{{\textbf{ANOVA $k_\text{best}$}}} & no & $\times$ & $\times$ & $\times$ & $\times$
            & $\times$
            \\
            & \num{10}    & $\times$ & $\times$ & $\times$  & $\times$ &          \\
            & \num{100}   & $\times$ & $\times$ & $\times$  &          &          \\
            & \num{1000}  & $\times$ & $\times$ &           &          &          \\
            & \num{10000} & $\times$ &          &           &          &          \\
            & \num{50000} & $\times$ &          &           &          &          \\

        \end{tabular}
        \vspace{6mm}
        \caption[Configurations for the preprocessing]{The $\times$ marks the used combinations of configurations for the preprocessing. The \emph{no}-label means that this preprocessing step were left out.}
        \label{tab:preprocessing}
    \end{center}
\end{table}

\FloatBarrier

\subsection{Results}
\label{cha:alternative-machine-configuration-results}
\begin{figure}
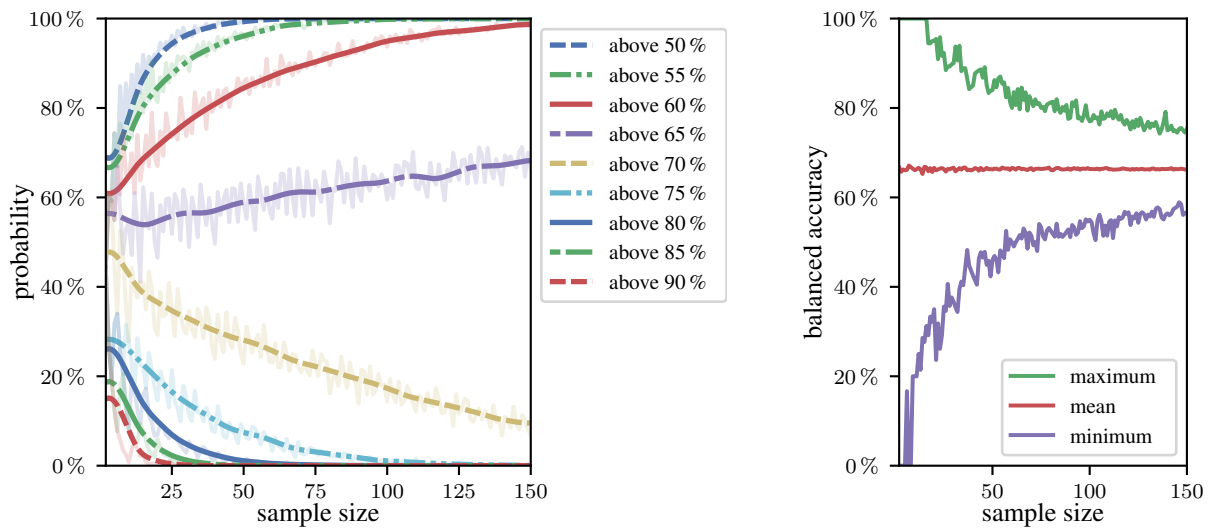

    \captionsetup[subfigure]{justification=justified,singlelinecheck=false}
    \begin{subfigure}[t]{0.61\textwidth}
        \input{images/no_PCA_no_selection_RandomForest_test_size_chances.pgf}
        \caption{Probabilities for Random Forest to yield an accuracy exceeding a certain threshold as a function of test sample size.}
    \end{subfigure}
    \hspace{3.0mm}
    \begin{subfigure}[t]{0.34\textwidth}
        \input{images/no_PCA_no_selection_RandomForest_test_size_stats.pgf}
        \caption{Minimum, maximum and mean results for the Random Forest as a function of test sample size.}
    \end{subfigure}
    \caption[Effects of varying test sample size. Random Forest; No preprocessing]{Results as a function of variable test set sizes with a fixed classifier. A \textbf{{Random Forest}} was trained with default parameters. ($n_\text{estimators}=\num{100}$)}
    \label{fig:no_PCA_no_selection_RandomForest}
\end{figure}

\begin{figure}
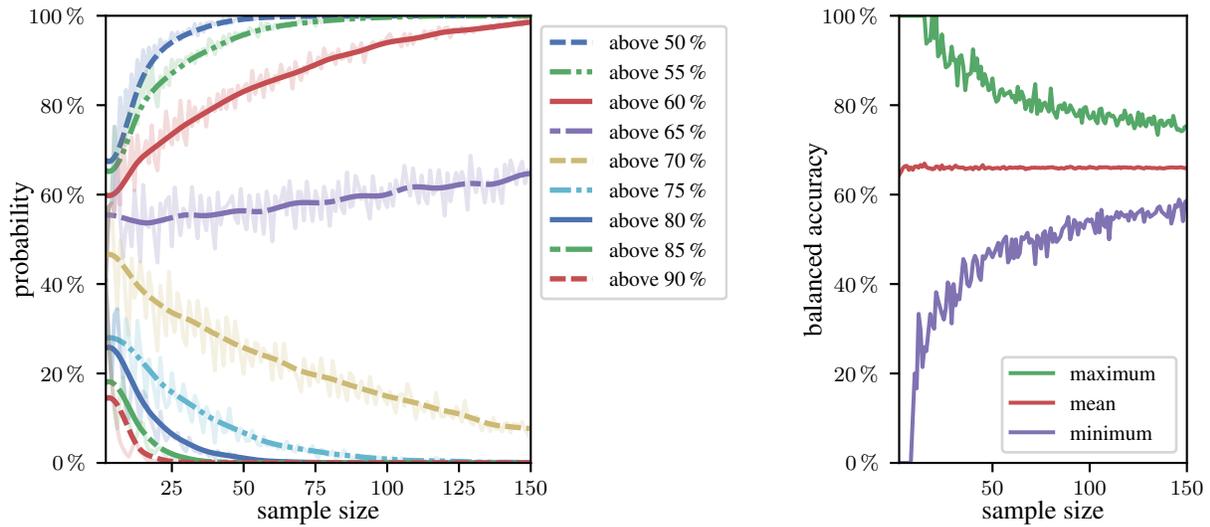

    \captionsetup[subfigure]{justification=justified,singlelinecheck=false}
    \begin{subfigure}[t]{0.61\textwidth}
        \input{images/no_PCA_no_selection_SVC_test_size_chances.pgf}
        \caption{Probabilities for SVM with an RBF kernel to yield an accuracy exceeding a certain threshold as a function of test sample size.}
    \end{subfigure}
    \hspace{3.0mm}
    \begin{subfigure}[t]{0.34\textwidth}
        \input{images/no_PCA_no_selection_SVC_test_size_stats.pgf}
        \caption{Minimum, maximum and mean results for the SVM with an RBF kernel as a function of test sample size.}
    \end{subfigure}
    \caption[Effects of varying test sample size. SVM (kernel = RBF); No preprocessing]{Results as a function of variable test set sizes with a fixed classifier. An \textbf{{SVM}} with an \textbf{{RBF kernel}} was trained with default parameters. ($C=\num{1.0}$; $\gamma=\sfrac{1}{n_\text{feature}}$)}
    \label{fig:no_PCA_no_selection_SVC}
\end{figure}

\begin{figure}
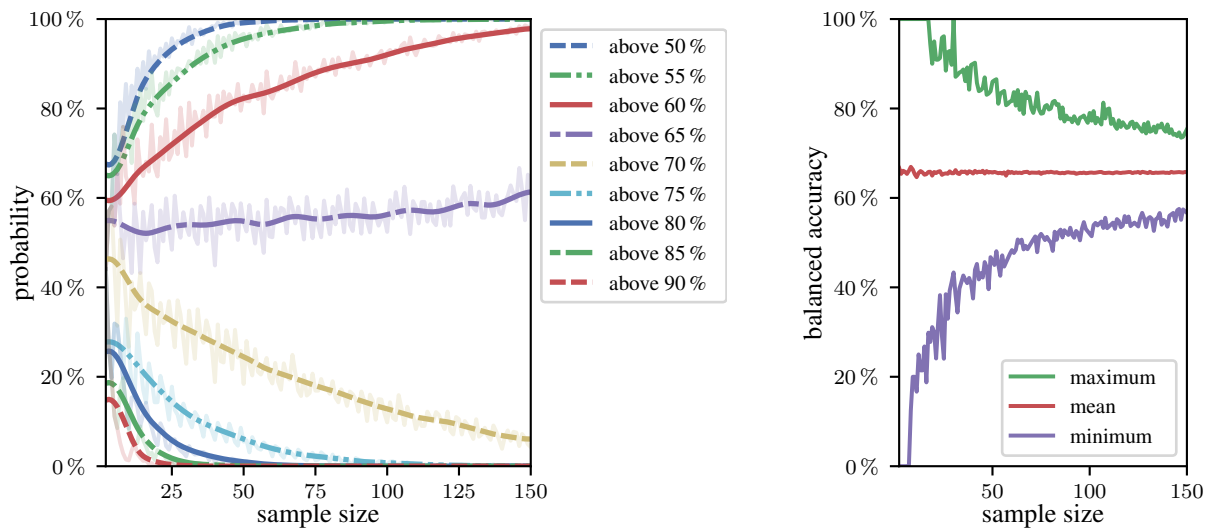

    \captionsetup[subfigure]{justification=justified,singlelinecheck=false}
    \begin{subfigure}[t]{0.61\textwidth}
        \input{images/no_PCA_no_selection_LinearSVC_test_size_chances.pgf}
        \caption{Probabilities for linear SVM to yield an accuracy exceeding a certain threshold as a function of test sample size.}
    \end{subfigure}
    \hspace{3.0mm}
    \begin{subfigure}[t]{0.34\textwidth}
        \input{images/no_PCA_no_selection_LinearSVC_test_size_stats.pgf}
        \caption{Minimum, maximum and mean results for the linear SVM as a function of test sample size.}
    \end{subfigure}
    \caption[Effects of varying test sample size. Linear SVM; No preprocessing]{Results as a function of variable test set sizes with a fixed classifier. A \textbf{{linear SVM}} was trained with default parameters. ($C=\num{1.0}$)}
    \label{fig:no_PCA_no_selection_LinearSVC}
\end{figure}

The results are comparable to the originally used configuration, an \ac{svm} with a linear kernel and no preprocessing (see Figure~\ref{fig:no_PCA_no_selection_RandomForest} to~\ref{fig:PCA_all_components_1000_best_selected_LinearSVC} and Table~\ref{tab:no_PCA_no_selection_RandomForest} to~\ref{tab:PCA_all_components_1000_best_selected_LinearSVC}). The results thus underline a generally valid character of the findings.
An outlier in the results can be found for a specific configuration, a \ac{pca} with 10 components and the \ac{svm} with an \ac{rbf} kernel (see Figure~\ref{fig:PCA_10_components_no_selection_SVC} and Table~\ref{tab:PCA_10_components_no_selection_SVC}). This result can be explained due to overfitting, the machine constantly returns one constant class as a prediction.
\FloatBarrier

\subsubsection{Graphical results representation}

\begin{figure}
    \captionsetup[subfigure]{justification=justified,singlelinecheck=false}
    \begin{subfigure}[t]{0.61\textwidth}
        \input{images/no_PCA_10_best_selected_RandomForest_test_size_chances.pgf}
        \caption{Probabilities for Random Forest to yield an accuracy exceeding a certain threshold as a function of test sample size.}
    \end{subfigure}
    \hspace{3.0mm}
    \begin{subfigure}[t]{0.34\textwidth}
        \input{images/no_PCA_10_best_selected_RandomForest_test_size_stats.pgf}
        \caption{Minimum, maximum and mean results for the Random Forest as a function of test sample size.}
    \end{subfigure}
    \caption[Effects of varying test sample size. Random Forest; Preprocessing: ANOVA feature selection ($k_\text{best} = \num{10}$)]{Results as a function of variable test set sizes with a fixed classifier. For \textbf{feature selection} an ANOVA was computed inside the the pipeline and the top \textbf{10 features} were taken based on the ANOVA F-values. Following, a \textbf{{Random Forest}} was trained with default parameters. ($n_\text{estimators}=\num{100}$)}
    \label{fig:no_PCA_10_best_selected_RandomForest}
\end{figure}

\begin{figure}
    \captionsetup[subfigure]{justification=justified,singlelinecheck=false}
    \begin{subfigure}[t]{0.61\textwidth}
        \input{images/no_PCA_10_best_selected_SVC_test_size_chances.pgf}
        \caption{Probabilities for SVM with an RBF kernel to yield an accuracy exceeding a certain threshold as a function of test sample size.}
    \end{subfigure}
    \hspace{3.0mm}
    \begin{subfigure}[t]{0.34\textwidth}
        \input{images/no_PCA_10_best_selected_SVC_test_size_stats.pgf}
        \caption{Minimum, maximum and mean results for the SVM with an RBF kernel as a function of test sample size.}
    \end{subfigure}
    \caption[Effects of varying test sample size. SVM (kernel = RBF); Preprocessing: ANOVA feature selection ($k_\text{best} = \num{10}$)]{Results as a function of variable test set sizes with a fixed classifier. For \textbf{feature selection} an ANOVA was computed inside the the pipeline and the top \textbf{10 features} were taken based on the ANOVA F-values. Following, an \textbf{{SVM}} with an \textbf{{RBF kernel}} was trained with default parameters. ($C=\num{1.0}$; $\gamma=\sfrac{1}{n_\text{feature}}$)}
    \label{fig:no_PCA_10_best_selected_SVC}
\end{figure}

\begin{figure}
    \captionsetup[subfigure]{justification=justified,singlelinecheck=false}
    \begin{subfigure}[t]{0.61\textwidth}
        \input{images/no_PCA_10_best_selected_LinearSVC_test_size_chances.pgf}
        \caption{Probabilities for linear SVM to yield an accuracy exceeding a certain threshold as a function of test sample size.}
    \end{subfigure}
    \hspace{3.0mm}
    \begin{subfigure}[t]{0.34\textwidth}
        \input{images/no_PCA_10_best_selected_LinearSVC_test_size_stats.pgf}
        \caption{Minimum, maximum and mean results for the linear SVM as a function of test sample size.}
    \end{subfigure}
    \caption[Effects of varying test sample size. Linear SVM; Preprocessing: ANOVA feature selection ($k_\text{best} = \num{10}$)]{Results as a function of variable test set sizes with a fixed classifier. For \textbf{feature selection} an ANOVA was computed inside the the pipeline and the top \textbf{10 features} were taken based on the ANOVA F-values. Following, a \textbf{{linear SVM}} was trained with default parameters. ($C=\num{1.0}$)}
    \label{fig:no_PCA_10_best_selected_LinearSVC}
\end{figure}

\begin{figure}
    \captionsetup[subfigure]{justification=justified,singlelinecheck=false}
    \begin{subfigure}[t]{0.61\textwidth}
        \input{images/no_PCA_100_best_selected_RandomForest_test_size_chances.pgf}
        \caption{Probabilities for Random Forest to yield an accuracy exceeding a certain threshold as a function of test sample size.}
    \end{subfigure}
    \hspace{3.0mm}
    \begin{subfigure}[t]{0.34\textwidth}
        \input{images/no_PCA_100_best_selected_RandomForest_test_size_stats.pgf}
        \caption{Minimum, maximum and mean results for the Random Forest as a function of test sample size.}
    \end{subfigure}
    \caption[Effects of varying test sample size. Random Forest; Preprocessing: ANOVA feature selection ($k_\text{best} = \num{100}$)]{Results as a function of variable test set sizes with a fixed classifier. For \textbf{feature selection} an ANOVA was computed inside the the pipeline and the top \textbf{100 features} were taken based on the ANOVA F-values. Following, a \textbf{{Random Forest}} was trained with default parameters. ($n_\text{estimators}=\num{100}$)}
    \label{fig:no_PCA_100_best_selected_RandomForest}
\end{figure}

\begin{figure}
    \captionsetup[subfigure]{justification=justified,singlelinecheck=false}
    \begin{subfigure}[t]{0.61\textwidth}
        \input{images/no_PCA_100_best_selected_SVC_test_size_chances.pgf}
        \caption{Probabilities for SVM with an RBF kernel to yield an accuracy exceeding a certain threshold as a function of test sample size.}
    \end{subfigure}
    \hspace{3.0mm}
    \begin{subfigure}[t]{0.34\textwidth}
        \input{images/no_PCA_100_best_selected_SVC_test_size_stats.pgf}
        \caption{Minimum, maximum and mean results for the SVM with an RBF kernel as a function of test sample size.}
    \end{subfigure}
    \caption[Effects of varying test sample size. SVM (kernel = RBF); Preprocessing: ANOVA feature selection ($k_\text{best} = \num{100}$)]{Results as a function of variable test set sizes with a fixed classifier. For \textbf{feature selection} an ANOVA was computed inside the the pipeline and the top \textbf{100 features} were taken based on the ANOVA F-values. Following, an \textbf{{SVM}} with an \textbf{{RBF kernel}} was trained with default parameters. ($C=\num{1.0}$; $\gamma=\sfrac{1}{n_\text{feature}}$)}
    \label{fig:no_PCA_100_best_selected_SVC}
\end{figure}

\begin{figure}
    \captionsetup[subfigure]{justification=justified,singlelinecheck=false}
    \begin{subfigure}[t]{0.61\textwidth}
        \input{images/no_PCA_100_best_selected_LinearSVC_test_size_chances.pgf}
        \caption{Probabilities for linear SVM to yield an accuracy exceeding a certain threshold as a function of test sample size.}
    \end{subfigure}
    \hspace{3.0mm}
    \begin{subfigure}[t]{0.34\textwidth}
        \input{images/no_PCA_100_best_selected_LinearSVC_test_size_stats.pgf}
        \caption{Minimum, maximum and mean results for the linear SVM as a function of test sample size.}
    \end{subfigure}
    \caption[Effects of varying test sample size. Linear SVM; Preprocessing: ANOVA feature selection ($k_\text{best} = \num{100}$)]{Results as a function of variable test set sizes with a fixed classifier. For \textbf{feature selection} an ANOVA was computed inside the the pipeline and the top \textbf{100 features} were taken based on the ANOVA F-values. Following, a \textbf{{linear SVM}} was trained with default parameters. ($C=\num{1.0}$)}
    \label{fig:no_PCA_100_best_selected_LinearSVC}
\end{figure}

\begin{figure}
    \captionsetup[subfigure]{justification=justified,singlelinecheck=false}
    \begin{subfigure}[t]{0.61\textwidth}
        \input{images/no_PCA_1000_best_selected_RandomForest_test_size_chances.pgf}
        \caption{Probabilities for Random Forest to yield an accuracy exceeding a certain threshold as a function of test sample size.}
    \end{subfigure}
    \hspace{3.0mm}
    \begin{subfigure}[t]{0.34\textwidth}
        \input{images/no_PCA_1000_best_selected_RandomForest_test_size_stats.pgf}
        \caption{Minimum, maximum and mean results for the Random Forest as a function of test sample size.}
    \end{subfigure}
    \caption[Effects of varying test sample size. Random Forest; Preprocessing: ANOVA feature selection ($k_\text{best} = \num{1000}$)]{Results as a function of variable test set sizes with a fixed classifier. For \textbf{feature selection} an ANOVA was computed inside the the pipeline and the top \textbf{1,000 features} were taken based on the ANOVA F-values. Following, a \textbf{{Random Forest}} was trained with default parameters. ($n_\text{estimators}=\num{100}$)}
    \label{fig:no_PCA_1000_best_selected_RandomForest}
\end{figure}

\begin{figure}
    \captionsetup[subfigure]{justification=justified,singlelinecheck=false}
    \begin{subfigure}[t]{0.61\textwidth}
        \input{images/no_PCA_1000_best_selected_SVC_test_size_chances.pgf}
        \caption{Probabilities for SVM with an RBF kernel to yield an accuracy exceeding a certain threshold as a function of test sample size.}
    \end{subfigure}
    \hspace{3.0mm}
    \begin{subfigure}[t]{0.34\textwidth}
        \input{images/no_PCA_1000_best_selected_SVC_test_size_stats.pgf}
        \caption{Minimum, maximum and mean results for the SVM with an RBF kernel as a function of test sample size.}
    \end{subfigure}
    \caption[Effects of varying test sample size. SVM (kernel = RBF); Preprocessing: ANOVA feature selection ($k_\text{best} = \num{1000}$)]{Results as a function of variable test set sizes with a fixed classifier. For \textbf{feature selection} an ANOVA was computed inside the the pipeline and the top \textbf{1,000 features} were taken based on the ANOVA F-values. Following, an \textbf{{SVM}} with an \textbf{{RBF kernel}} was trained with default parameters. ($C=\num{1.0}$; $\gamma=\sfrac{1}{n_\text{feature}}$)}
    \label{fig:no_PCA_1000_best_selected_SVC}
\end{figure}

\begin{figure}
    \captionsetup[subfigure]{justification=justified,singlelinecheck=false}
    \begin{subfigure}[t]{0.61\textwidth}
        \input{images/no_PCA_1000_best_selected_LinearSVC_test_size_chances.pgf}
        \caption{Probabilities for linear SVM to yield an accuracy exceeding a certain threshold as a function of test sample size.}
    \end{subfigure}
    \hspace{3.0mm}
    \begin{subfigure}[t]{0.34\textwidth}
        \input{images/no_PCA_1000_best_selected_LinearSVC_test_size_stats.pgf}
        \caption{Minimum, maximum and mean results for the linear SVM as a function of test sample size.}
    \end{subfigure}
    \caption[Effects of varying test sample size. Linear SVM; Preprocessing: ANOVA feature selection ($k_\text{best} = \num{1000}$)]{Results as a function of variable test set sizes with a fixed classifier. For \textbf{feature selection} an ANOVA was computed inside the the pipeline and the top \textbf{1,000 features} were taken based on the ANOVA F-values. Following, a \textbf{{linear SVM}} was trained with default parameters. ($C=\num{1.0}$)}
    \label{fig:no_PCA_1000_best_selected_LinearSVC}
\end{figure}

\begin{figure}
    \captionsetup[subfigure]{justification=justified,singlelinecheck=false}
    \begin{subfigure}[t]{0.61\textwidth}
        \input{images/no_PCA_10000_best_selected_RandomForest_test_size_chances.pgf}
        \caption{Probabilities for Random Forest to yield an accuracy exceeding a certain threshold as a function of test sample size.}
    \end{subfigure}
    \hspace{3.0mm}
    \begin{subfigure}[t]{0.34\textwidth}
        \input{images/no_PCA_10000_best_selected_RandomForest_test_size_stats.pgf}
        \caption{Minimum, maximum and mean results for the Random Forest as a function of test sample size.}
    \end{subfigure}
    \caption[Effects of varying test sample size. Random Forest; Preprocessing: ANOVA feature selection ($k_\text{best} = \num{10000}$)]{Results as a function of variable test set sizes with a fixed classifier. For \textbf{feature selection} an ANOVA was computed inside the the pipeline and the top \textbf{10,000 features} were taken based on the ANOVA F-values. Following, a \textbf{{Random Forest}} was trained with default parameters. ($n_\text{estimators}=\num{100}$)}
    \label{fig:no_PCA_10000_best_selected_RandomForest}
\end{figure}

\begin{figure}
    \captionsetup[subfigure]{justification=justified,singlelinecheck=false}
    \begin{subfigure}[t]{0.61\textwidth}
        \input{images/no_PCA_10000_best_selected_SVC_test_size_chances.pgf}
        \caption{Probabilities for SVM with an RBF kernel to yield an accuracy exceeding a certain threshold as a function of test sample size.}
    \end{subfigure}
    \hspace{3.0mm}
    \begin{subfigure}[t]{0.34\textwidth}
        \input{images/no_PCA_10000_best_selected_SVC_test_size_stats.pgf}
        \caption{Minimum, maximum and mean results for the SVM with an RBF kernel as a function of test sample size.}
    \end{subfigure}
    \caption[Effects of varying test sample size. SVM (kernel = RBF); Preprocessing: ANOVA feature selection ($k_\text{best} = \num{10000}$)]{Results as a function of variable test set sizes with a fixed classifier. For \textbf{feature selection} an ANOVA was computed inside the the pipeline and the top \textbf{10,000 features} were taken based on the ANOVA F-values. Following, an \textbf{{SVM}} with an \textbf{{RBF kernel}} was trained with default parameters. ($C=\num{1.0}$; $\gamma=\sfrac{1}{n_\text{feature}}$)}
    \label{fig:no_PCA_10000_best_selected_SVC}
\end{figure}

\begin{figure}
    \captionsetup[subfigure]{justification=justified,singlelinecheck=false}
    \begin{subfigure}[t]{0.61\textwidth}
        \input{images/no_PCA_10000_best_selected_LinearSVC_test_size_chances.pgf}
        \caption{Probabilities for linear SVM to yield an accuracy exceeding a certain threshold as a function of test sample size.}
    \end{subfigure}
    \hspace{3.0mm}
    \begin{subfigure}[t]{0.34\textwidth}
        \input{images/no_PCA_10000_best_selected_LinearSVC_test_size_stats.pgf}
        \caption{Minimum, maximum and mean results for the linear SVM as a function of test sample size.}
    \end{subfigure}
    \caption[Effects of varying test sample size. Linear SVM; Preprocessing: ANOVA feature selection ($k_\text{best} = \num{10000}$)]{Results as a function of variable test set sizes with a fixed classifier. For \textbf{feature selection} an ANOVA was computed inside the the pipeline and the top \textbf{10,000 features} were taken based on the ANOVA F-values. Following, a \textbf{{linear SVM}} was trained with default parameters. ($C=\num{1.0}$)}
    \label{fig:no_PCA_10000_best_selected_LinearSVC}
\end{figure}

\begin{figure}
    \captionsetup[subfigure]{justification=justified,singlelinecheck=false}
    \begin{subfigure}[t]{0.61\textwidth}
        \input{images/no_PCA_50000_best_selected_RandomForest_test_size_chances.pgf}
        \caption{Probabilities for Random Forest to yield an accuracy exceeding a certain threshold as a function of test sample size.}
    \end{subfigure}
    \hspace{3.0mm}
    \begin{subfigure}[t]{0.34\textwidth}
        \input{images/no_PCA_50000_best_selected_RandomForest_test_size_stats.pgf}
        \caption{Minimum, maximum and mean results for the Random Forest as a function of test sample size.}
    \end{subfigure}
    \caption[Effects of varying test sample size. Random Forest; Preprocessing: ANOVA feature selection ($k_\text{best} = \num{50000}$)]{Results as a function of variable test set sizes with a fixed classifier. For \textbf{feature selection} an ANOVA was computed inside the the pipeline and the top \textbf{50,000 features} were taken based on the ANOVA F-values. Following, a \textbf{{Random Forest}} was trained with default parameters. ($n_\text{estimators}=\num{100}$)}
    \label{fig:no_PCA_50000_best_selected_RandomForest}
\end{figure}

\begin{figure}
    \captionsetup[subfigure]{justification=justified,singlelinecheck=false}
    \begin{subfigure}[t]{0.61\textwidth}
        \input{images/no_PCA_50000_best_selected_SVC_test_size_chances.pgf}
        \caption{Probabilities for SVM with an RBF kernel to yield an accuracy exceeding a certain threshold as a function of test sample size.}
    \end{subfigure}
    \hspace{3.0mm}
    \begin{subfigure}[t]{0.34\textwidth}
        \input{images/no_PCA_50000_best_selected_SVC_test_size_stats.pgf}
        \caption{Minimum, maximum and mean results for the SVM with an RBF kernel as a function of test sample size.}
    \end{subfigure}
    \caption[Effects of varying test sample size. SVM (kernel = RBF); Preprocessing: ANOVA feature selection ($k_\text{best} = \num{50000}$)]{Results as a function of variable test set sizes with a fixed classifier. For \textbf{feature selection} an ANOVA was computed inside the the pipeline and the top \textbf{50,000 features} were taken based on the ANOVA F-values. Following, an \textbf{{SVM}} with an \textbf{{RBF kernel}} was trained with default parameters. ($C=\num{1.0}$; $\gamma=\sfrac{1}{n_\text{feature}}$)}
    \label{fig:no_PCA_50000_best_selected_SVC}
\end{figure}

\begin{figure}
    \captionsetup[subfigure]{justification=justified,singlelinecheck=false}
    \begin{subfigure}[t]{0.61\textwidth}
        \input{images/no_PCA_50000_best_selected_LinearSVC_test_size_chances.pgf}
        \caption{Probabilities for linear SVM to yield an accuracy exceeding a certain threshold as a function of test sample size.}
    \end{subfigure}
    \hspace{3.0mm}
    \begin{subfigure}[t]{0.34\textwidth}
        \input{images/no_PCA_50000_best_selected_LinearSVC_test_size_stats.pgf}
        \caption{Minimum, maximum and mean results for the linear SVM as a function of test sample size.}
    \end{subfigure}
    \caption[Effects of varying test sample size. Linear SVM; Preprocessing: ANOVA feature selection ($k_\text{best} = \num{50000}$)]{Results as a function of variable test set sizes with a fixed classifier. For \textbf{feature selection} an ANOVA was computed inside the the pipeline and the top \textbf{50,000 features} were taken based on the ANOVA F-values. Following, a \textbf{{linear SVM}} was trained with default parameters. ($C=\num{1.0}$)}
    \label{fig:no_PCA_50000_best_selected_LinearSVC}
\end{figure}

\begin{figure}
    \captionsetup[subfigure]{justification=justified,singlelinecheck=false}
    \begin{subfigure}[t]{0.61\textwidth}
        \input{images/PCA_10_components_no_selection_RandomForest_test_size_chances.pgf}
        \caption{Probabilities for Random Forest to yield an accuracy exceeding a certain threshold as a function of test sample size.}
    \end{subfigure}
    \hspace{3.0mm}
    \begin{subfigure}[t]{0.34\textwidth}
        \input{images/PCA_10_components_no_selection_RandomForest_test_size_stats.pgf}
        \caption{Minimum, maximum and mean results for the Random Forest as a function of test sample size.}
    \end{subfigure}
    \caption[Effects of varying test sample size. Random Forest; Preprocessing: PCA ($n_\text{components} = \num{10}$)]{Results as a function of variable test set sizes with a fixed classifier. To reduce the dimensionality of the feature space a \textbf{PCA} was performed and \textbf{10 components} were retained. Following, a \textbf{{Random Forest}} was trained with default parameters. ($n_\text{estimators}=\num{100}$)}
    \label{fig:PCA_10_components_no_selection_RandomForest}
\end{figure}

\begin{figure}
    \captionsetup[subfigure]{justification=justified,singlelinecheck=false}
    \begin{subfigure}[t]{0.61\textwidth}
        \input{images/PCA_10_components_no_selection_SVC_test_size_chances.pgf}
        \caption{Probabilities for SVM with an RBF kernel to yield an accuracy exceeding a certain threshold as a function of test sample size.}
    \end{subfigure}
    \hspace{3.0mm}
    \begin{subfigure}[t]{0.34\textwidth}
        \input{images/PCA_10_components_no_selection_SVC_test_size_stats.pgf}
        \caption{Minimum, maximum and mean results for the SVM with an RBF kernel as a function of test sample size.}
    \end{subfigure}
    \caption[Effects of varying test sample size. SVM (kernel = RBF); Preprocessing: PCA ($n_\text{components} = \num{10}$)]{Results as a function of variable test set sizes with a fixed classifier. To reduce the dimensionality of the feature space a \textbf{PCA} was performed and \textbf{10 components} were retained. Following, an \textbf{{SVM}} with an \textbf{{RBF kernel}} was trained with default parameters. ($C=\num{1.0}$; $\gamma=\sfrac{1}{n_\text{feature}}$)}
    \label{fig:PCA_10_components_no_selection_SVC}
\end{figure}

\begin{figure}
    \captionsetup[subfigure]{justification=justified,singlelinecheck=false}
    \begin{subfigure}[t]{0.61\textwidth}
        \input{images/PCA_10_components_no_selection_LinearSVC_test_size_chances.pgf}
        \caption{Probabilities for linear SVM to yield an accuracy exceeding a certain threshold as a function of test sample size.}
    \end{subfigure}
    \hspace{3.0mm}
    \begin{subfigure}[t]{0.34\textwidth}
        \input{images/PCA_10_components_no_selection_LinearSVC_test_size_stats.pgf}
        \caption{Minimum, maximum and mean results for the linear SVM as a function of test sample size.}
    \end{subfigure}
    \caption[Effects of varying test sample size. Linear SVM; Preprocessing: PCA ($n_\text{components} = \num{10}$)]{Results as a function of variable test set sizes with a fixed classifier. To reduce the dimensionality of the feature space a \textbf{PCA} was performed and \textbf{10 components} were retained. Following, a \textbf{{linear SVM}} was trained with default parameters. ($C=\num{1.0}$)}
    \label{fig:PCA_10_components_no_selection_LinearSVC}
\end{figure}

\begin{figure}
    \captionsetup[subfigure]{justification=justified,singlelinecheck=false}
    \begin{subfigure}[t]{0.61\textwidth}
        \input{images/PCA_50_components_no_selection_RandomForest_test_size_chances.pgf}
        \caption{Probabilities for Random Forest to yield an accuracy exceeding a certain threshold as a function of test sample size.}
    \end{subfigure}
    \hspace{3.0mm}
    \begin{subfigure}[t]{0.34\textwidth}
        \input{images/PCA_50_components_no_selection_RandomForest_test_size_stats.pgf}
        \caption{Minimum, maximum and mean results for the Random Forest as a function of test sample size.}
    \end{subfigure}
    \caption[Effects of varying test sample size. Random Forest; Preprocessing: PCA ($n_\text{components} = \num{50}$)]{Results as a function of variable test set sizes with a fixed classifier. To reduce the dimensionality of the feature space a \textbf{PCA} was performed and \textbf{50 components} were retained. Following, a \textbf{{Random Forest}} was trained with default parameters. ($n_\text{estimators}=\num{100}$)}
    \label{fig:PCA_50_components_no_selection_RandomForest}
\end{figure}

\begin{figure}
    \captionsetup[subfigure]{justification=justified,singlelinecheck=false}
    \begin{subfigure}[t]{0.61\textwidth}
        \input{images/PCA_50_components_no_selection_SVC_test_size_chances.pgf}
        \caption{Probabilities for SVM with an RBF kernel to yield an accuracy exceeding a certain threshold as a function of test sample size.}
    \end{subfigure}
    \hspace{3.0mm}
    \begin{subfigure}[t]{0.34\textwidth}
        \input{images/PCA_50_components_no_selection_SVC_test_size_stats.pgf}
        \caption{Minimum, maximum and mean results for the SVM with an RBF kernel as a function of test sample size.}
    \end{subfigure}
    \caption[Effects of varying test sample size. SVM (kernel = RBF); Preprocessing: PCA ($n_\text{components} = \num{50}$)]{Results as a function of variable test set sizes with a fixed classifier. To reduce the dimensionality of the feature space a \textbf{PCA} was performed and \textbf{50 components} were retained. Following, an \textbf{{SVM}} with an \textbf{{RBF kernel}} was trained with default parameters. ($C=\num{1.0}$; $\gamma=\sfrac{1}{n_\text{feature}}$)}
    \label{fig:PCA_50_components_no_selection_SVC}
\end{figure}

\begin{figure}
    \captionsetup[subfigure]{justification=justified,singlelinecheck=false}
    \begin{subfigure}[t]{0.61\textwidth}
        \input{images/PCA_50_components_no_selection_LinearSVC_test_size_chances.pgf}
        \caption{Probabilities for linear SVM to yield an accuracy exceeding a certain threshold as a function of test sample size.}
    \end{subfigure}
    \hspace{3.0mm}
    \begin{subfigure}[t]{0.34\textwidth}
        \input{images/PCA_50_components_no_selection_LinearSVC_test_size_stats.pgf}
        \caption{Minimum, maximum and mean results for the linear SVM as a function of test sample size.}
    \end{subfigure}
    \caption[Effects of varying test sample size. Linear SVM; Preprocessing: PCA ($n_\text{components} = \num{50}$)]{Results as a function of variable test set sizes with a fixed classifier. To reduce the dimensionality of the feature space a \textbf{PCA} was performed and \textbf{50 components} were retained. Following, a \textbf{{linear SVM}} was trained with default parameters. ($C=\num{1.0}$)}
    \label{fig:PCA_50_components_no_selection_LinearSVC}
\end{figure}

\begin{figure}
    \captionsetup[subfigure]{justification=justified,singlelinecheck=false}
    \begin{subfigure}[t]{0.61\textwidth}
        \input{images/PCA_50_components_10_best_selected_RandomForest_test_size_chances.pgf}
        \caption{Probabilities for Random Forest to yield an accuracy exceeding a certain threshold as a function of test sample size.}
    \end{subfigure}
    \hspace{3.0mm}
    \begin{subfigure}[t]{0.34\textwidth}
        \input{images/PCA_50_components_10_best_selected_RandomForest_test_size_stats.pgf}
        \caption{Minimum, maximum and mean results for the Random Forest as a function of test sample size.}
    \end{subfigure}
    \caption[Effects of varying test sample size. Random Forest; Preprocessing: PCA ($n_\text{components} = \num{50}$); ANOVA feature selection ($k_\text{best} = \num{10}$)]{Results as a function of variable test set sizes with a fixed classifier. To reduce the dimensionality of the feature space a \textbf{PCA} was performed and \textbf{50 components} were retained. For \textbf{feature selection} an ANOVA was computed inside the the pipeline and the top \textbf{10 features} were taken based on the ANOVA F-values. Following, a \textbf{{Random Forest}} was trained with default parameters. ($n_\text{estimators}=\num{100}$)}
    \label{fig:PCA_50_components_10_best_selected_RandomForest}
\end{figure}

\begin{figure}
    \captionsetup[subfigure]{justification=justified,singlelinecheck=false}
    \begin{subfigure}[t]{0.61\textwidth}
        \input{images/PCA_50_components_10_best_selected_SVC_test_size_chances.pgf}
        \caption{Probabilities for SVM with an RBF kernel to yield an accuracy exceeding a certain threshold as a function of test sample size.}
    \end{subfigure}
    \hspace{3.0mm}
    \begin{subfigure}[t]{0.34\textwidth}
        \input{images/PCA_50_components_10_best_selected_SVC_test_size_stats.pgf}
        \caption{Minimum, maximum and mean results for the SVM with an RBF kernel as a function of test sample size.}
    \end{subfigure}
    \caption[Effects of varying test sample size. SVM (kernel = RBF); Preprocessing: PCA ($n_\text{components} = \num{50}$); ANOVA feature selection ($k_\text{best} = \num{10}$)]{Results as a function of variable test set sizes with a fixed classifier. To reduce the dimensionality of the feature space a \textbf{PCA} was performed and \textbf{50 components} were retained. For \textbf{feature selection} an ANOVA was computed inside the the pipeline and the top \textbf{10 features} were taken based on the ANOVA F-values. Following, an \textbf{{SVM}} with an \textbf{{RBF kernel}} was trained with default parameters. ($C=\num{1.0}$; $\gamma=\sfrac{1}{n_\text{feature}}$)}
    \label{fig:PCA_50_components_10_best_selected_SVC}
\end{figure}

\begin{figure}
    \captionsetup[subfigure]{justification=justified,singlelinecheck=false}
    \begin{subfigure}[t]{0.61\textwidth}
        \input{images/PCA_50_components_10_best_selected_LinearSVC_test_size_chances.pgf}
        \caption{Probabilities for linear SVM to yield an accuracy exceeding a certain threshold as a function of test sample size.}
    \end{subfigure}
    \hspace{3.0mm}
    \begin{subfigure}[t]{0.34\textwidth}
        \input{images/PCA_50_components_10_best_selected_LinearSVC_test_size_stats.pgf}
        \caption{Minimum, maximum and mean results for the linear SVM as a function of test sample size.}
    \end{subfigure}
    \caption[Effects of varying test sample size. Linear SVM; Preprocessing: PCA ($n_\text{components} = \num{50}$); ANOVA feature selection ($k_\text{best} = \num{10}$)]{Results as a function of variable test set sizes with a fixed classifier. To reduce the dimensionality of the feature space a \textbf{PCA} was performed and \textbf{50 components} were retained. For \textbf{feature selection} an ANOVA was computed inside the the pipeline and the top \textbf{10 features} were taken based on the ANOVA F-values. Following, a \textbf{{linear SVM}} was trained with default parameters. ($C=\num{1.0}$)}
    \label{fig:PCA_50_components_10_best_selected_LinearSVC}
\end{figure}

\begin{figure}
    \captionsetup[subfigure]{justification=justified,singlelinecheck=false}
    \begin{subfigure}[t]{0.61\textwidth}
        \input{images/PCA_500_components_no_selection_RandomForest_test_size_chances.pgf}
        \caption{Probabilities for Random Forest to yield an accuracy exceeding a certain threshold as a function of test sample size.}
    \end{subfigure}
    \hspace{3.0mm}
    \begin{subfigure}[t]{0.34\textwidth}
        \input{images/PCA_500_components_no_selection_RandomForest_test_size_stats.pgf}
        \caption{Minimum, maximum and mean results for the Random Forest as a function of test sample size.}
    \end{subfigure}
    \caption[Effects of varying test sample size. Random Forest; Preprocessing: PCA ($n_\text{components} = \num{500}$)]{Results as a function of variable test set sizes with a fixed classifier. To reduce the dimensionality of the feature space a \textbf{PCA} was performed and \textbf{500 components} were retained. Following, a \textbf{{Random Forest}} was trained with default parameters. ($n_\text{estimators}=\num{100}$)}
    \label{fig:PCA_500_components_no_selection_RandomForest}
\end{figure}

\begin{figure}
    \captionsetup[subfigure]{justification=justified,singlelinecheck=false}
    \begin{subfigure}[t]{0.61\textwidth}
        \input{images/PCA_500_components_no_selection_SVC_test_size_chances.pgf}
        \caption{Probabilities for SVM with an RBF kernel to yield an accuracy exceeding a certain threshold as a function of test sample size.}
    \end{subfigure}
    \hspace{3.0mm}
    \begin{subfigure}[t]{0.34\textwidth}
        \input{images/PCA_500_components_no_selection_SVC_test_size_stats.pgf}
        \caption{Minimum, maximum and mean results for the SVM with an RBF kernel as a function of test sample size.}
    \end{subfigure}
    \caption[Effects of varying test sample size. SVM (kernel = RBF); Preprocessing: PCA ($n_\text{components} = \num{500}$)]{Results as a function of variable test set sizes with a fixed classifier. To reduce the dimensionality of the feature space a \textbf{PCA} was performed and \textbf{500 components} were retained. Following, an \textbf{{SVM}} with an \textbf{{RBF kernel}} was trained with default parameters. ($C=\num{1.0}$; $\gamma=\sfrac{1}{n_\text{feature}}$)}
    \label{fig:PCA_500_components_no_selection_SVC}
\end{figure}

\begin{figure}
    \captionsetup[subfigure]{justification=justified,singlelinecheck=false}
    \begin{subfigure}[t]{0.61\textwidth}
        \input{images/PCA_500_components_no_selection_LinearSVC_test_size_chances.pgf}
        \caption{Probabilities for linear SVM to yield an accuracy exceeding a certain threshold as a function of test sample size.}
    \end{subfigure}
    \hspace{3.0mm}
    \begin{subfigure}[t]{0.34\textwidth}
        \input{images/PCA_500_components_no_selection_LinearSVC_test_size_stats.pgf}
        \caption{Minimum, maximum and mean results for the linear SVM as a function of test sample size.}
    \end{subfigure}
    \caption[Effects of varying test sample size. Linear SVM; Preprocessing: PCA ($n_\text{components} = \num{500}$)]{Results as a function of variable test set sizes with a fixed classifier. To reduce the dimensionality of the feature space a \textbf{PCA} was performed and \textbf{500 components} were retained. Following, a \textbf{{linear SVM}} was trained with default parameters. ($C=\num{1.0}$)}
    \label{fig:PCA_500_components_no_selection_LinearSVC}
\end{figure}

\begin{figure}
    \captionsetup[subfigure]{justification=justified,singlelinecheck=false}
    \begin{subfigure}[t]{0.61\textwidth}
        \input{images/PCA_500_components_10_best_selected_RandomForest_test_size_chances.pgf}
        \caption{Probabilities for Random Forest to yield an accuracy exceeding a certain threshold as a function of test sample size.}
    \end{subfigure}
    \hspace{3.0mm}
    \begin{subfigure}[t]{0.34\textwidth}
        \input{images/PCA_500_components_10_best_selected_RandomForest_test_size_stats.pgf}
        \caption{Minimum, maximum and mean results for the Random Forest as a function of test sample size.}
    \end{subfigure}
    \caption[Effects of varying test sample size. Random Forest; Preprocessing: PCA ($n_\text{components} = \num{500}$); ANOVA feature selection ($k_\text{best} = \num{10}$)]{Results as a function of variable test set sizes with a fixed classifier. To reduce the dimensionality of the feature space a \textbf{PCA} was performed and \textbf{500 components} were retained. For \textbf{feature selection} an ANOVA was computed inside the the pipeline and the top \textbf{10 features} were taken based on the ANOVA F-values. Following, a \textbf{{Random Forest}} was trained with default parameters. ($n_\text{estimators}=\num{100}$)}
    \label{fig:PCA_500_components_10_best_selected_RandomForest}
\end{figure}

\begin{figure}
    \captionsetup[subfigure]{justification=justified,singlelinecheck=false}
    \begin{subfigure}[t]{0.61\textwidth}
        \input{images/PCA_500_components_10_best_selected_SVC_test_size_chances.pgf}
        \caption{Probabilities for SVM with an RBF kernel to yield an accuracy exceeding a certain threshold as a function of test sample size.}
    \end{subfigure}
    \hspace{3.0mm}
    \begin{subfigure}[t]{0.34\textwidth}
        \input{images/PCA_500_components_10_best_selected_SVC_test_size_stats.pgf}
        \caption{Minimum, maximum and mean results for the SVM with an RBF kernel as a function of test sample size.}
    \end{subfigure}
    \caption[Effects of varying test sample size. SVM (kernel = RBF); Preprocessing: PCA ($n_\text{components} = \num{500}$); ANOVA feature selection ($k_\text{best} = \num{10}$)]{Results as a function of variable test set sizes with a fixed classifier. To reduce the dimensionality of the feature space a \textbf{PCA} was performed and \textbf{500 components} were retained. For \textbf{feature selection} an ANOVA was computed inside the the pipeline and the top \textbf{10 features} were taken based on the ANOVA F-values. Following, an \textbf{{SVM}} with an \textbf{{RBF kernel}} was trained with default parameters. ($C=\num{1.0}$; $\gamma=\sfrac{1}{n_\text{feature}}$)}
    \label{fig:PCA_500_components_10_best_selected_SVC}
\end{figure}

\begin{figure}
    \captionsetup[subfigure]{justification=justified,singlelinecheck=false}
    \begin{subfigure}[t]{0.61\textwidth}
        \input{images/PCA_500_components_10_best_selected_LinearSVC_test_size_chances.pgf}
        \caption{Probabilities for linear SVM to yield an accuracy exceeding a certain threshold as a function of test sample size.}
    \end{subfigure}
    \hspace{3.0mm}
    \begin{subfigure}[t]{0.34\textwidth}
        \input{images/PCA_500_components_10_best_selected_LinearSVC_test_size_stats.pgf}
        \caption{Minimum, maximum and mean results for the linear SVM as a function of test sample size.}
    \end{subfigure}
    \caption[Effects of varying test sample size. Linear SVM; Preprocessing: PCA ($n_\text{components} = \num{500}$); ANOVA feature selection ($k_\text{best} = \num{10}$)]{Results as a function of variable test set sizes with a fixed classifier. To reduce the dimensionality of the feature space a \textbf{PCA} was performed and \textbf{500 components} were retained. For \textbf{feature selection} an ANOVA was computed inside the the pipeline and the top \textbf{10 features} were taken based on the ANOVA F-values. Following, a \textbf{{linear SVM}} was trained with default parameters. ($C=\num{1.0}$)}
    \label{fig:PCA_500_components_10_best_selected_LinearSVC}
\end{figure}

\begin{figure}
    \captionsetup[subfigure]{justification=justified,singlelinecheck=false}
    \begin{subfigure}[t]{0.61\textwidth}
        \input{images/PCA_500_components_100_best_selected_RandomForest_test_size_chances.pgf}
        \caption{Probabilities for Random Forest to yield an accuracy exceeding a certain threshold as a function of test sample size.}
    \end{subfigure}
    \hspace{3.0mm}
    \begin{subfigure}[t]{0.34\textwidth}
        \input{images/PCA_500_components_100_best_selected_RandomForest_test_size_stats.pgf}
        \caption{Minimum, maximum and mean results for the Random Forest as a function of test sample size.}
    \end{subfigure}
    \caption[Effects of varying test sample size. Random Forest; Preprocessing: PCA ($n_\text{components} = \num{500}$); ANOVA feature selection ($k_\text{best} = \num{100}$)]{Results as a function of variable test set sizes with a fixed classifier. To reduce the dimensionality of the feature space a \textbf{PCA} was performed and \textbf{500 components} were retained. For \textbf{feature selection} an ANOVA was computed inside the the pipeline and the top \textbf{100 features} were taken based on the ANOVA F-values. Following, a \textbf{{Random Forest}} was trained with default parameters. ($n_\text{estimators}=\num{100}$)}
    \label{fig:PCA_500_components_100_best_selected_RandomForest}
\end{figure}

\begin{figure}
    \captionsetup[subfigure]{justification=justified,singlelinecheck=false}
    \begin{subfigure}[t]{0.61\textwidth}
        \input{images/PCA_500_components_100_best_selected_SVC_test_size_chances.pgf}
        \caption{Probabilities for SVM with an RBF kernel to yield an accuracy exceeding a certain threshold as a function of test sample size.}
    \end{subfigure}
    \hspace{3.0mm}
    \begin{subfigure}[t]{0.34\textwidth}
        \input{images/PCA_500_components_100_best_selected_SVC_test_size_stats.pgf}
        \caption{Minimum, maximum and mean results for the SVM with an RBF kernel as a function of test sample size.}
    \end{subfigure}
    \caption[Effects of varying test sample size. SVM (kernel = RBF); Preprocessing: PCA ($n_\text{components} = \num{500}$); ANOVA feature selection ($k_\text{best} = \num{100}$)]{Results as a function of variable test set sizes with a fixed classifier. To reduce the dimensionality of the feature space a \textbf{PCA} was performed and \textbf{500 components} were retained. For \textbf{feature selection} an ANOVA was computed inside the the pipeline and the top \textbf{100 features} were taken based on the ANOVA F-values. Following, an \textbf{{SVM}} with an \textbf{{RBF kernel}} was trained with default parameters. ($C=\num{1.0}$; $\gamma=\sfrac{1}{n_\text{feature}}$)}
    \label{fig:PCA_500_components_100_best_selected_SVC}
\end{figure}

\begin{figure}
    \captionsetup[subfigure]{justification=justified,singlelinecheck=false}
    \begin{subfigure}[t]{0.61\textwidth}
        \input{images/PCA_500_components_100_best_selected_LinearSVC_test_size_chances.pgf}
        \caption{Probabilities for linear SVM to yield an accuracy exceeding a certain threshold as a function of test sample size.}
    \end{subfigure}
    \hspace{3.0mm}
    \begin{subfigure}[t]{0.34\textwidth}
        \input{images/PCA_500_components_100_best_selected_LinearSVC_test_size_stats.pgf}
        \caption{Minimum, maximum and mean results for the linear SVM as a function of test sample size.}
    \end{subfigure}
    \caption[Effects of varying test sample size. Linear SVM; Preprocessing: PCA ($n_\text{components} = \num{500}$); ANOVA feature selection ($k_\text{best} = \num{100}$)]{Results as a function of variable test set sizes with a fixed classifier. To reduce the dimensionality of the feature space a \textbf{PCA} was performed and \textbf{500 components} were retained. For \textbf{feature selection} an ANOVA was computed inside the the pipeline and the top \textbf{100 features} were taken based on the ANOVA F-values. Following, a \textbf{{linear SVM}} was trained with default parameters. ($C=\num{1.0}$)}
    \label{fig:PCA_500_components_100_best_selected_LinearSVC}
\end{figure}

\begin{figure}
    \captionsetup[subfigure]{justification=justified,singlelinecheck=false}
    \begin{subfigure}[t]{0.61\textwidth}
        \input{images/PCA_all_components_no_selection_RandomForest_test_size_chances.pgf}
        \caption{Probabilities for Random Forest to yield an accuracy exceeding a certain threshold as a function of test sample size.}
    \end{subfigure}
    \hspace{3.0mm}
    \begin{subfigure}[t]{0.34\textwidth}
        \input{images/PCA_all_components_no_selection_RandomForest_test_size_stats.pgf}
        \caption{Minimum, maximum and mean results for the Random Forest as a function of test sample size.}
    \end{subfigure}
    \caption[Effects of varying test sample size. Random Forest; Preprocessing: PCA ($n_\text{components} = \text{all}$)]{Results as a function of variable test set sizes with a fixed classifier. To reduce the dimensionality of the feature space a \textbf{PCA} was performed and \textbf{all components} were retained. Following, a \textbf{{Random Forest}} was trained with default parameters. ($n_\text{estimators}=\num{100}$)}
    \label{fig:PCA_all_components_no_selection_RandomForest}
\end{figure}

\begin{figure}
    \captionsetup[subfigure]{justification=justified,singlelinecheck=false}
    \begin{subfigure}[t]{0.61\textwidth}
        \input{images/PCA_all_components_no_selection_SVC_test_size_chances.pgf}
        \caption{Probabilities for SVM with an RBF kernel to yield an accuracy exceeding a certain threshold as a function of test sample size.}
    \end{subfigure}
    \hspace{3.0mm}
    \begin{subfigure}[t]{0.34\textwidth}
        \input{images/PCA_all_components_no_selection_SVC_test_size_stats.pgf}
        \caption{Minimum, maximum and mean results for the SVM with an RBF kernel as a function of test sample size.}
    \end{subfigure}
    \caption[Effects of varying test sample size. SVM (kernel = RBF); Preprocessing: PCA ($n_\text{components} = \text{all}$)]{Results as a function of variable test set sizes with a fixed classifier. To reduce the dimensionality of the feature space a \textbf{PCA} was performed and \textbf{all components} were retained. Following, an \textbf{{SVM}} with an \textbf{{RBF kernel}} was trained with default parameters. ($C=\num{1.0}$; $\gamma=\sfrac{1}{n_\text{feature}}$)}
    \label{fig:PCA_all_components_no_selection_SVC}
\end{figure}

\begin{figure}
    \captionsetup[subfigure]{justification=justified,singlelinecheck=false}
    \begin{subfigure}[t]{0.61\textwidth}
        \input{images/PCA_all_components_no_selection_LinearSVC_test_size_chances.pgf}
        \caption{Probabilities for linear SVM to yield an accuracy exceeding a certain threshold as a function of test sample size.}
    \end{subfigure}
    \hspace{3.0mm}
    \begin{subfigure}[t]{0.34\textwidth}
        \input{images/PCA_all_components_no_selection_LinearSVC_test_size_stats.pgf}
        \caption{Minimum, maximum and mean results for the linear SVM as a function of test sample size.}
    \end{subfigure}
    \caption[Effects of varying test sample size. Linear SVM; Preprocessing: PCA ($n_\text{components} = \text{all}$)]{Results as a function of variable test set sizes with a fixed classifier. To reduce the dimensionality of the feature space a \textbf{PCA} was performed and \textbf{all components} were retained. Following, a \textbf{{linear SVM}} was trained with default parameters. ($C=\num{1.0}$)}
    \label{fig:PCA_all_components_no_selection_LinearSVC}
\end{figure}

\begin{figure}
    \captionsetup[subfigure]{justification=justified,singlelinecheck=false}
    \begin{subfigure}[t]{0.61\textwidth}
        \input{images/PCA_all_components_10_best_selected_RandomForest_test_size_chances.pgf}
        \caption{Probabilities for Random Forest to yield an accuracy exceeding a certain threshold as a function of test sample size.}
    \end{subfigure}
    \hspace{3.0mm}
    \begin{subfigure}[t]{0.34\textwidth}
        \input{images/PCA_all_components_10_best_selected_RandomForest_test_size_stats.pgf}
        \caption{Minimum, maximum and mean results for the Random Forest as a function of test sample size.}
    \end{subfigure}
    \caption[Effects of varying test sample size. Random Forest; Preprocessing: PCA ($n_\text{components} = \text{all}$); ANOVA feature selection ($k_\text{best} = \num{10}$)]{Results as a function of variable test set sizes with a fixed classifier. To reduce the dimensionality of the feature space a \textbf{PCA} was performed and \textbf{all components} were retained. For \textbf{feature selection} an ANOVA was computed inside the the pipeline and the top \textbf{10 features} were taken based on the ANOVA F-values. Following, a \textbf{{Random Forest}} was trained with default parameters. ($n_\text{estimators}=\num{100}$)}
    \label{fig:PCA_all_components_10_best_selected_RandomForest}
\end{figure}

\begin{figure}
    \captionsetup[subfigure]{justification=justified,singlelinecheck=false}
    \begin{subfigure}[t]{0.61\textwidth}
        \input{images/PCA_all_components_10_best_selected_SVC_test_size_chances.pgf}
        \caption{Probabilities for SVM with an RBF kernel to yield an accuracy exceeding a certain threshold as a function of test sample size.}
    \end{subfigure}
    \hspace{3.0mm}
    \begin{subfigure}[t]{0.34\textwidth}
        \input{images/PCA_all_components_10_best_selected_SVC_test_size_stats.pgf}
        \caption{Minimum, maximum and mean results for the SVM with an RBF kernel as a function of test sample size.}
    \end{subfigure}
    \caption[Effects of varying test sample size. SVM (kernel = RBF); Preprocessing: PCA ($n_\text{components} = \text{all}$); ANOVA feature selection ($k_\text{best} = \num{10}$)]{Results as a function of variable test set sizes with a fixed classifier. To reduce the dimensionality of the feature space a \textbf{PCA} was performed and \textbf{all components} were retained. For \textbf{feature selection} an ANOVA was computed inside the the pipeline and the top \textbf{10 features} were taken based on the ANOVA F-values. Following, an \textbf{{SVM}} with an \textbf{{RBF kernel}} was trained with default parameters. ($C=\num{1.0}$; $\gamma=\sfrac{1}{n_\text{feature}}$)}
    \label{fig:PCA_all_components_10_best_selected_SVC}
\end{figure}

\begin{figure}
    \captionsetup[subfigure]{justification=justified,singlelinecheck=false}
    \begin{subfigure}[t]{0.61\textwidth}
        \input{images/PCA_all_components_10_best_selected_LinearSVC_test_size_chances.pgf}
        \caption{Probabilities for linear SVM to yield an accuracy exceeding a certain threshold as a function of test sample size.}
    \end{subfigure}
    \hspace{3.0mm}
    \begin{subfigure}[t]{0.34\textwidth}
        \input{images/PCA_all_components_10_best_selected_LinearSVC_test_size_stats.pgf}
        \caption{Minimum, maximum and mean results for the linear SVM as a function of test sample size.}
    \end{subfigure}
    \caption[Effects of varying test sample size. Linear SVM; Preprocessing: PCA ($n_\text{components} = \text{all}$); ANOVA feature selection ($k_\text{best} = \num{10}$)]{Results as a function of variable test set sizes with a fixed classifier. To reduce the dimensionality of the feature space a \textbf{PCA} was performed and \textbf{all components} were retained. For \textbf{feature selection} an ANOVA was computed inside the the pipeline and the top \textbf{10 features} were taken based on the ANOVA F-values. Following, a \textbf{{linear SVM}} was trained with default parameters. ($C=\num{1.0}$)}
    \label{fig:PCA_all_components_10_best_selected_LinearSVC}
\end{figure}

\begin{figure}
    \captionsetup[subfigure]{justification=justified,singlelinecheck=false}
    \begin{subfigure}[t]{0.61\textwidth}
        \input{images/PCA_all_components_100_best_selected_RandomForest_test_size_chances.pgf}
        \caption{Probabilities for Random Forest to yield an accuracy exceeding a certain threshold as a function of test sample size.}
    \end{subfigure}
    \hspace{3.0mm}
    \begin{subfigure}[t]{0.34\textwidth}
        \input{images/PCA_all_components_100_best_selected_RandomForest_test_size_stats.pgf}
        \caption{Minimum, maximum and mean results for the Random Forest as a function of test sample size.}
    \end{subfigure}
    \caption[Effects of varying test sample size. Random Forest; Preprocessing: PCA ($n_\text{components} = \text{all}$); ANOVA feature selection ($k_\text{best} = \num{100}$)]{Results as a function of variable test set sizes with a fixed classifier. To reduce the dimensionality of the feature space a \textbf{PCA} was performed and \textbf{all components} were retained. For \textbf{feature selection} an ANOVA was computed inside the the pipeline and the top \textbf{100 features} were taken based on the ANOVA F-values. Following, a \textbf{{Random Forest}} was trained with default parameters. ($n_\text{estimators}=\num{100}$)}
    \label{fig:PCA_all_components_100_best_selected_RandomForest}
\end{figure}

\begin{figure}
    \captionsetup[subfigure]{justification=justified,singlelinecheck=false}
    \begin{subfigure}[t]{0.61\textwidth}
        \input{images/PCA_all_components_100_best_selected_SVC_test_size_chances.pgf}
        \caption{Probabilities for SVM with an RBF kernel to yield an accuracy exceeding a certain threshold as a function of test sample size.}
    \end{subfigure}
    \hspace{3.0mm}
    \begin{subfigure}[t]{0.34\textwidth}
        \input{images/PCA_all_components_100_best_selected_SVC_test_size_stats.pgf}
        \caption{Minimum, maximum and mean results for the SVM with an RBF kernel as a function of test sample size.}
    \end{subfigure}
    \caption[Effects of varying test sample size. SVM (kernel = RBF); Preprocessing: PCA ($n_\text{components} = \text{all}$); ANOVA feature selection ($k_\text{best} = \num{100}$)]{Results as a function of variable test set sizes with a fixed classifier. To reduce the dimensionality of the feature space a \textbf{PCA} was performed and \textbf{all components} were retained. For \textbf{feature selection} an ANOVA was computed inside the the pipeline and the top \textbf{100 features} were taken based on the ANOVA F-values. Following, an \textbf{{SVM}} with an \textbf{{RBF kernel}} was trained with default parameters. ($C=\num{1.0}$; $\gamma=\sfrac{1}{n_\text{feature}}$)}
    \label{fig:PCA_all_components_100_best_selected_SVC}
\end{figure}

\begin{figure}
    \captionsetup[subfigure]{justification=justified,singlelinecheck=false}
    \begin{subfigure}[t]{0.61\textwidth}
        \input{images/PCA_all_components_100_best_selected_LinearSVC_test_size_chances.pgf}
        \caption{Probabilities for linear SVM to yield an accuracy exceeding a certain threshold as a function of test sample size.}
    \end{subfigure}
    \hspace{3.0mm}
    \begin{subfigure}[t]{0.34\textwidth}
        \input{images/PCA_all_components_100_best_selected_LinearSVC_test_size_stats.pgf}
        \caption{Minimum, maximum and mean results for the linear SVM as a function of test sample size.}
    \end{subfigure}
    \caption[Effects of varying test sample size. Linear SVM; Preprocessing: PCA ($n_\text{components} = \text{all}$); ANOVA feature selection ($k_\text{best} = \num{100}$)]{Results as a function of variable test set sizes with a fixed classifier. To reduce the dimensionality of the feature space a \textbf{PCA} was performed and \textbf{all components} were retained. For \textbf{feature selection} an ANOVA was computed inside the the pipeline and the top \textbf{100 features} were taken based on the ANOVA F-values. Following, a \textbf{{linear SVM}} was trained with default parameters. ($C=\num{1.0}$)}
    \label{fig:PCA_all_components_100_best_selected_LinearSVC}
\end{figure}

\begin{figure}
    \captionsetup[subfigure]{justification=justified,singlelinecheck=false}
    \begin{subfigure}[t]{0.61\textwidth}
        \input{images/PCA_all_components_1000_best_selected_RandomForest_test_size_chances.pgf}
        \caption{Probabilities for Random Forest to yield an accuracy exceeding a certain threshold as a function of test sample size.}
    \end{subfigure}
    \hspace{3.0mm}
    \begin{subfigure}[t]{0.34\textwidth}
        \input{images/PCA_all_components_1000_best_selected_RandomForest_test_size_stats.pgf}
        \caption{Minimum, maximum and mean results for the Random Forest as a function of test sample size.}
    \end{subfigure}
    \caption[Effects of varying test sample size. Random Forest; Preprocessing: PCA ($n_\text{components} = \text{all}$); ANOVA feature selection ($k_\text{best} = \num{1000}$)]{Results as a function of variable test set sizes with a fixed classifier. To reduce the dimensionality of the feature space a \textbf{PCA} was performed and \textbf{all components} were retained. For \textbf{feature selection} an ANOVA was computed inside the the pipeline and the top \textbf{1,000 features} were taken based on the ANOVA F-values. Following, a \textbf{{Random Forest}} was trained with default parameters. ($n_\text{estimators}=\num{100}$)}
    \label{fig:PCA_all_components_1000_best_selected_RandomForest}
\end{figure}

\begin{figure}
    \captionsetup[subfigure]{justification=justified,singlelinecheck=false}
    \begin{subfigure}[t]{0.61\textwidth}
        \input{images/PCA_all_components_1000_best_selected_SVC_test_size_chances.pgf}
        \caption{Probabilities for SVM with an RBF kernel to yield an accuracy exceeding a certain threshold as a function of test sample size.}
    \end{subfigure}
    \hspace{3.0mm}
    \begin{subfigure}[t]{0.34\textwidth}
        \input{images/PCA_all_components_1000_best_selected_SVC_test_size_stats.pgf}
        \caption{Minimum, maximum and mean results for the SVM with an RBF kernel as a function of test sample size.}
    \end{subfigure}
    \caption[Effects of varying test sample size. SVM (kernel = RBF); Preprocessing: PCA ($n_\text{components} = \text{all}$); ANOVA feature selection ($k_\text{best} = \num{1000}$)]{Results as a function of variable test set sizes with a fixed classifier. To reduce the dimensionality of the feature space a \textbf{PCA} was performed and \textbf{all components} were retained. For \textbf{feature selection} an ANOVA was computed inside the the pipeline and the top \textbf{1,000 features} were taken based on the ANOVA F-values. Following, an \textbf{{SVM}} with an \textbf{{RBF kernel}} was trained with default parameters. ($C=\num{1.0}$; $\gamma=\sfrac{1}{n_\text{feature}}$)}
    \label{fig:PCA_all_components_1000_best_selected_SVC}
\end{figure}

\begin{figure}
    \captionsetup[subfigure]{justification=justified,singlelinecheck=false}
    \begin{subfigure}[t]{0.61\textwidth}
        \input{images/PCA_all_components_1000_best_selected_LinearSVC_test_size_chances.pgf}
        \caption{Probabilities for linear SVM to yield an accuracy exceeding a certain threshold as a function of test sample size.}
    \end{subfigure}
    \hspace{3.0mm}
    \begin{subfigure}[t]{0.34\textwidth}
        \input{images/PCA_all_components_1000_best_selected_LinearSVC_test_size_stats.pgf}
        \caption{Minimum, maximum and mean results for the linear SVM as a function of test sample size.}
    \end{subfigure}
    \caption[Effects of varying test sample size. Linear SVM; Preprocessing: PCA ($n_\text{components} = \text{all}$); ANOVA feature selection ($k_\text{best} = \num{1000}$)]{Results as a function of variable test set sizes with a fixed classifier. To reduce the dimensionality of the feature space a \textbf{PCA} was performed and \textbf{all components} were retained. For \textbf{feature selection} an ANOVA was computed inside the the pipeline and the top \textbf{1,000 features} were taken based on the ANOVA F-values. Following, a \textbf{{linear SVM}} was trained with default parameters. ($C=\num{1.0}$)}
    \label{fig:PCA_all_components_1000_best_selected_LinearSVC}
\end{figure}

\FloatBarrier

\subsubsection{Tabular results representation}
\input{supplements_alternative_machines_tabular.tex}

    \FloatBarrier
    
    \section{Abbreviations}
    \begin{acronym}
    \acro{mdd}[MDD]{major depressive disorder}
    \acro{hc}[HC]{healthy control}
    \acro{pac}[PAC]{Predictive Analytics Competition}
    \acro{ml}[ML]{machine learning}
    \acro{mri}[MRI]{magnetic resonance imaging}
    \acro{adhd}[ADHD]{attention deficit hyperactivity disorder}
    \acro{loocv}[LOOCV]{leave-one-out cross-validation}
    \acro{svm}[SVM]{support-vector machine}
    \acroplural{svm}[SVMs]{support-vector machines}
    \acro{tiv}[TIV]{total intracranial volume}
    \acro{pca}[PCA]{principal-component analysis}
    \acro{rbf}[RBF]{radial basis function}
\end{acronym}

\end{document}